\documentclass[sigconf,nonacm]{acmart}
\usepackage{graphicx}
\usepackage{booktabs,tabularx}
\usepackage{enumitem}
\microtypesetup{expansion=false}
\usepackage{amsmath}
\usepackage{tikz}
\usepackage{soul}
\usepackage{pgfplots}
\pgfplotsset{compat=1.18}
\usetikzlibrary{positioning,arrows.meta,calc,decorations.pathreplacing}

\definecolor{openred}{HTML}{A8452F}
\definecolor{slate}{HTML}{1F3B54}
\definecolor{softgrey}{HTML}{EEF1F4}
\definecolor{accent}{HTML}{4A6B45}
\definecolor{mockblue}{HTML}{2B5E9C}

\newcommand{\stem}[1]{\textit{#1}}
\newcommand{\vreq}[1]{\textbf{R#1}}

\newcommand{\tabref}[1]{Table~\ref{#1}}
\newcommand{\figref}[1]{Figure~\ref{#1}}

\newcommand{\secref}[1]{Section~\ref{#1}}

\newcommand{\appref}[1]{Appendix~\ref{#1}}

\begin{document}

\title{Time Machine Experiments: Using Historically-Bounded AI for Inquiry into the Human Mind}

\author{Hiromu Yakura}
\authornote{These authors contributed equally to this work.}
\author{Robin Schimmelpfennig}
\authornotemark[1]
\author{Ezequiel Lopez-Lopez}
\authornotemark[1]
\additionalaffiliation{%
  \institution{TUD Dresden University of Technology}
  \city{Dresden}
  \country{Germany}
}
\author{Alejandro H. Artiles}
\author{Levin Brinkmann}
\affiliation{%
  \institution{Max Planck Institute for Human Development}
  \city{Berlin}
  \country{Germany}
}

\makeatletter
\let\firstrowaddresses\addresses
\gdef\addresses{}
\makeatother

\author{Jean-Fran\c{c}ois Bonnefon}
\affiliation{%
  \institution{Toulouse School of Economics}
  \city{Toulouse}
  \country{France}
}

\author{Azim Shariff}
\affiliation{%
  \institution{The University of British Columbia}
  \city{Vancouver}
  \country{Canada}
}

\author{Iyad Rahwan}
\affiliation{%
  \institution{Max Planck Institute for Human Development}
  \city{Berlin}
  \country{Germany}
}

\makeatletter
\let\secondrowaddresses\addresses
\makeatother

\makeatletter

\let\acm@mkauthors@iii\@mkauthors@iii

\def\@mkauthors@iii{%
  \global\setbox\mktitle@bx=\vbox{%
    \noindent
    \unvbox\mktitle@bx
    \par\medskip

    \centering

    {\@authorfont
      Hiromu Yakura\footnotemark[1],
      Robin Schimmelpfennig\footnotemark[1],
      Ezequiel Lopez-Lopez\footnotemark[1]\footnotemark[2],
      \par
      Alejandro Hernandez Artiles,
      Levin Brinkmann
      \par
    }

    \smallskip

    {\@affiliationfont
      Max Planck Institute for Human Development,
      Berlin, Germany
      \par
    }

    \medskip
  }%

  \begingroup
    \let\addresses\secondrowaddresses
    \num@authorgroups=3\relax
    \acm@mkauthors@iii
  \endgroup
}

\makeatother

\begin{abstract}
Can interacting with someone from 1930 who has no knowledge of what happened after, influence a person's perception of the past? People reason about the present against a picture of the past without observing it. The past is reconstructed from memory and testimony, but this reconstruction has been filtered through everything that happened since. Historically-bounded large language models (LLMs) make that past available for interaction. As a proof-of-concept for the impact of interacting with historical minds, we ran a preregistered randomized experiment ($N=240$), where participants interacted with an LLM trained on pre-1930 text. The interaction reduced the illusion of moral decline, the tendency to view the past as more moral than the present, compared to the contemporary-model control. This \textit{Time Machine Experiment} paradigm informs new forms of interactive experiments, where temporal knowledge boundaries become experimental variables, and expands the realm of \emph{science fiction science}, which turns thought experiments into actual experiments.
\end{abstract}


\begin{teaserfigure}
 \includegraphics[width=\textwidth]{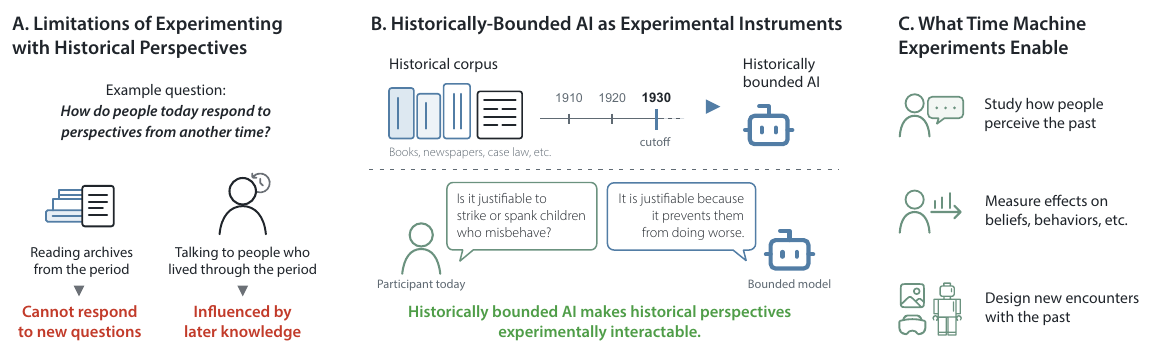}
  \caption{\textbf{Conceptual overview of Time Machine Experiments.}
  (A) Existing approaches provide limited ways to experimentally engage with historical perspectives: archival records cannot respond to new questions, while people who lived through a historical period are subsequently shaped by knowledge of later events.
  (B) Historically bounded models trained on corpora ending at a specified temporal cutoff make such perspectives experimentally interactable, allowing contemporary participants to ask, probe, and respond to questions that were not necessarily recorded in the historical archive.
  (C) The Time Machine Experiment paradigm encompasses studying how people perceive the past, measuring effects on beliefs, reflection, or behavior, and designing new forms of encounters with historical perspectives through interaction.}
  \Description{Three-panel conceptual diagram. The left panel (A) shows that existing approaches either use fixed historical records that cannot answer new questions or people whose perspectives include later knowledge. The center panel (B) states that a historically bounded AI trained on material before a temporal cutoff enables responsive interaction with contemporary participants. The right panel (C) illustrates that this enables studying perceptions of the past, measuring effects on people, and designing new historical encounters.}
  \label{fig:teaser}
\end{teaserfigure}

\maketitle
\renewcommand{\shortauthors}{Yakura, Schimmelpfennig, Lopez-Lopez, et al.}

\section{Introduction}
\label{sec:intro}

Advances in computing technology have repeatedly expanded the scales at which humans can be studied, from modeling biological processes at the molecular level~\cite{jumper2021highly} to tracing collective behavior through population-scale data~\cite{lazer2009computational}.
Human-computer interaction (HCI) research has contributed to this advancement by turning emerging computing technologies into methodological instruments, for example, using virtual reality to experimentally manipulate social situations and examine their effects~\cite{blascovich2002immersive,miller2021synchrony} or using interactive probes to investigate everyday human experience~\cite{hutchinson2003technology,rosner2009reflections}.
Large language models (LLMs) offer HCI another such methodological opportunity, enabling researchers to model, probe, and intervene in human cognition~\cite{danry2023dont,binz2025foundation,xu2025productive}.
More specifically, HCI can design frameworks and interfaces that treat LLMs as research instruments, systems whose interactions with humans can surface and reshape people's perceptions, assumptions, and values~\cite{costello2024durably,kobis2025delegation,danry2025deceptive}.
Here, we focus on one specific aspect of these opportunities, namely that LLMs trained on vast collections of human-produced text can provide direct, interactive access to patterns embedded in cultural and historical corpora in a way that directly accessing the human-produced text can not~\cite{argyle2023out,gurnee2024language,buttrick2024studying,lu2025cultural}.

There now exist language models trained exclusively on historical text corpora (e.g., before 1930~\cite{talkie2026model}), in part to examine whether models can rediscover or predict ideas, innovations, or societal changes that emerged only after their temporal cutoff.
These historically-bounded models open a methodological possibility for probing human behavior and cognition in an area where our current approaches to answering research questions are constrained.
We cannot talk to a historical book, we cannot talk to dead people, and we cannot literally travel back in time to conduct experiments in past societies~\cite{atari2023historical}.
We can talk to people alive today who have witnessed their past.
But they have also witnessed the time up until today, which means that their evaluation and perception of the past is biased.
Historically-bounded language models can create experimentally accessible approximations of such situations.

Building on this perspective, we introduce the \textit{Time Machine Experiment} (\figref{fig:teaser}), a methodological framework for constructing cross-temporal interactions with historically-bounded AI models to examine how people respond to change in how humans perceive, translate, and reason across time.
Our conceptualization of the Time Machine Experiment is situated across several HCI traditions, such as design fiction and speculative design, which construct alternative or counterfactual technological scenarios to provoke reflection on existing social assumptions~\cite{blythe2017research,elsden2017on}.
Our suggested approach uses methodological advancements made possible via LLMs to turn science-fiction-like scenarios into objects of scientific inquiry.
Thereby, we complement the \textit{Science Fiction Science} approach~\cite{rahwan2025science}, which complements the qualitative insights of science fiction writers with quantitative evidence about human responses to alternative assumptions.

To show the promise and feasibility of the Time Machine Experiment, we conducted a preregistered online experiment ($N=240$) centered on the `Illusion of Moral Decline', which describes people’s tendency to perceive past societies as more moral than contemporary ones~\cite{mastroianni2023illusion}.
To the extent that we can evaluate the accuracy perception via survey data, it is probably not true that past societies were more moral.
Instead, people overestimate the moral values of people in the past across several moral categories.
Based on this view and building on past work, we formulated the following research question: Can interacting with a model trained on data up until the 1930s change people's (biased) perception of moral values in that time?
The results show that compared to the control groups (i.e., interacting with a current frontier model: \texttt{gpt-5.5}), interaction with the historically-bounded model prompted substantially greater self-reported reflection and changes in participants' views of the past, thereby mitigating the illusion of moral decline.
These shifts in perception and reflection provide proof of concept that otherwise inaccessible cross-temporal interactions can be made experimentally tractable for studying and influencing human behaviors.
Based on these findings, we discuss the opportunities, limitations, and broader design space of Time Machine Experiment as a new methodological approach for HCI and behavioral research.

\section{Related work}
\label{sec:rw}

\subsection{Interactive systems as instruments for studying and influencing humans}

Interactive systems have long served in HCI and adjacent behavioral sciences both as objects of evaluation and as methodological apparatuses for human inquiry and intervention, such as virtual~\cite{blascovich2002immersive,miller2021synchrony} and interactive probes~\cite{hutchinson2003technology,rosner2009reflections}.
In particular, the tradition of prototyping allows people to gain first-hand access to existing or not-yet-realized conditions through active engagement with designed artifacts~\cite{buchenau2000experience}, thereby deliberately embodying tensions that surface otherwise tacit assumptions among participants~\cite{boer2012provotypes}.
This logic extends to speculative approaches in HCI, which construct consequential but unrealized situations for participants to inhabit and reflect upon~\cite{elsden2017on}.
For example, this approach has been used to challenge the conventional future orientation of speculation by treating `temporality' itself as a malleable design material to ~\cite{soro2019designing,kozubaev2020expanding}.
Taken together, we can see how HCI can generate novel knowledge about humans by deliberately designing interaction conditions under which otherwise difficult-to-access forms of perception, reasoning, and reflection become observable and open to inquiry.

HCI has also designed interactive systems as interventions to reshape human cognition and behavior.
Embodied perspective, taking in virtual reality, for example, has been shown to improve communication and reflection in interpersonal conflict~\cite{yong2024change,hirzle2026experiencing}, while reflective interventions can help people revisit and reinterpret their own experiences~\cite{arakawa2020inward,bentvelzen2022revisiting}. More recently, human--AI interaction has been explored as an intervention in domains ranging from coaching and self-reflection~\cite{arakawa2024coaching} to critical reasoning~\cite{danry2023dont}.
These papers show how HCI can extend the methodological role of interactive systems from understanding human cognition and behavior to actively shaping how people perceive, reflect, and reason.

Time Machine Experiment builds on these studies by treating temporality as a property of the interaction itself.
Prior work has already explored forms of experiential \textit{time travel}, for example, using embodied virtual reality (VR) to let participants experience past, present, and possible future stages of climate change and thereby influence subsequent attitudes and behavior~\cite{pi2025embodied}.
Yet such approaches primarily construct authored representations of another time.
As we discuss in the next section, recent advances in language modeling now make it possible to create interactive counterparts whose available knowledge is itself bounded by a historical period.
This enables a new paradigm for understanding and intervening in humans by enabling people to engage in cross-temporal interactions.

\subsection{LLM trained on historical corpus}

The continuous mainstreaming of AI research, making training pipelines, base models, and reinforcement learning algorithms more available for researchers in the social sciences, has contributed to the emerging idea of building historically-bounded language models.
For example, TimeCapsuleLLM~\cite{grigorian2025time}, TypewriterLM~\cite{xiaoxi2026pretraining}, GPT-1900~\cite{hla2026machina}, and Talkie~\cite{talkie2026model} are the models trained from scratch using temporally bounded corpora to avoid contamination from modern knowledge. 
This development has attracted interest not only for the evaluation of temporal generalization and forecasting, but also for behavioral science.
Varnum et al. discussed the potential of using historical LLMs as simulated participants that might approximate psychological responses of populations no longer available for direct study~\cite{varnum2024historical}.

Among these models, Talkie is currently one of the largest historically-bounded models, with 13 billion parameters pretrained on 260 billion tokens of English-language text published before 1930, consisting of books, newspapers, periodicals, scientific journals, patents, and case law~\cite{talkie2026model}.
Because essentially all such material originates in physical documents, its training corpus depends heavily on text recognition.
Nevertheless, its corpus construction processes explicitly avoid modern VLM-based recognition models, because they can hallucinate contemporary facts into historical text, and instead involved conventional OCR.
Talkie is also notable for combining relatively large-scale historical pretraining with post-training specifically designed to support conversational interaction.
The resulting model is therefore not merely prompted to imitate an earlier period, but is constructed around an information environment intentionally bounded in time.

This distinction is methodologically consequential.
Prompting a contemporary LLM to behave as if it had an earlier knowledge cutoff does not reliably remove later knowledge, especially when future knowledge is only causally related to a query~\cite{gao2025prompts}.
Instead, historically-bounded models enable a qualitatively different interaction condition, in which the temporal constraint is grounded in the model's training, instead than enacted through role-playing.
This aspect is key for the science fiction science approach~\cite{rahwan2025science}, in terms of turning otherwise science-fictional scenarios---such as interacting with an interlocutor whose knowledge is bounded to a century ago---into experimentally tractable objects of inquiry.
To the best of our knowledge, this paper is among the first attempts to articulate the design space of using historically-bounded language models for studying contemporary humans and to demonstrate a working instantiation through a proof-of-concept experiment.

\section{Time Machine Experiments}
\label{sec:instrument}

A \emph{Time Machine Experiment} uses interaction with an AI model whose information is bounded at a specified historical point as an intervention on people living now. Participants encounter the historically-bounded model, and the experiment can measure how they interact with it and how the interaction affects their beliefs, judgments, reflection, or behavior relative to a comparison condition. The contemporary participant is therefore the target of the intervention, while the model provides an experimental counterpart whose temporal boundary (cutoff) can be documented and tested. In the terms of science fiction science, contact with another time is the thought experiment, and the historically-bounded AI is its experimental surrogate~\cite{rahwan2025science}.

The paradigm is defined by the informational boundary of the deployed AI model rather than by a particular architecture or modality. 
A Time Machine Experiment may involve text, speech, images, audiovisual environments, embodied agents, or combinations of these, provided that the whole interactive system built on the model can be evaluated for information introduced after the stated boundary. Participants may question the system, predict how it will respond, confront it with present-day ideas, or explore a generated environment. 
Researchers can then examine how they interact in the experiment system and whether the interaction revises a belief about the past, makes assumptions rooted in the present visible, prompts reflection, or changes a subsequent decision. 
The case study in \secref{sec:case} presents a first and deliberately minimal instantiation, using dialogue and prediction tasks with a text-based language model to examine the influence of the interaction on perceived moral decline and reflective insight.

\subsection{Why Time Machine Experiments?}
\label{sec:instrument:why}

Behavioral scientists have developed several ways to expose participants to situations they cannot encounter directly. Experimental vignettes provide standardized and manipulable scenarios, while virtual environments can combine experimental control with richer forms of engagement~\cite{aguinis2014best,blascovich2002immersive,bombari2015studying}. These methods afford different degrees of interactivity, but common implementations constrain the encounter to scenarios and responses specified or mediated by researchers~\cite{bombari2015studying}. This constraint matters because reciprocal interaction is not psychologically equivalent to observing a fixed representation, as an interactive counterpart can respond to participants' own questions and allow unanticipated lines of inquiry to shape the encounter~\cite{redcay2019using}.

Studying the past adds the problem of temporal perspective. Archival materials provide historically situated evidence, but they are fixed and reflect selective processes of production and preservation~\cite{inwood2020selection,trouillot1995silencing}. Living witnesses can respond to new questions, but their recollections and retrospective judgments may be shaped by subsequent experiences and outcome knowledge~\cite{roese2012hindsight}.
They therefore do not provide an experimental condition in which information acquired after the historical period is held absent.
Contemporary AI can generate responsive historical simulations, but instructing a model to adopt a period persona does not reliably exclude leaked knowledge acquired after that period~\cite{gao2025prompts}.

Drawing these methodological concerns together, we identify three properties that historically-bounded AI can uniquely combine within a single experimental instrument: it can respond to questions the researcher did not anticipate, its information environment can have a documentable temporal boundary, and its encounter with participants can be assigned and compared experimentally. 
These properties do not make the system a historical person or establish what a historical population believed. They make a model conditioned by a bounded historical record available as a responsive experimental counterpart. 
Whether a particular system warrants this description, and what can be inferred from interacting with it, depends on the validity requirements we state next.

\subsection{Conditions for valid inference}
\label{sec:instrument:validity}

These unique properties inherently constrain the implementation of Time Machine Experiments, especially to support the claims that the model participants meet is genuinely bounded at its stated date and that what happens to participants can be attributed to that boundary. 
Four things must hold for both halves of that claim: the corpus must support the interpretation placed on the system, the boundary must hold in the system as deployed and not only in the base model, what participants encounter must be attributable to the model rather than produced by the design, and the comparison condition must support the causal attribution drawn from it.

\tabref{tab:validity} states each requirement as a question, the way the inference fails if the question goes unanswered, and the evidence that discharges it.
Although these are not new kinds of validity~\cite{shadish2002experimental}, two of them have no counterpart in ordinary experimental practice: an ordinary manipulation has no informational boundary that can leak (\vreq{2}), and an ordinary control condition can be built to differ in the manipulated factor alone, whereas no available model differs from a historically-bounded one in its cutoff alone (\vreq{4}).
What our own case study offers against each is reported in \secref{sec:case:materials} and \appref{app:materials}.
 
\begin{table*}[t]
\caption{Four requirements for a Time Machine Experiment: the question each asks, how the inference fails if it is left unanswered, and the evidence that discharges it. The requirements are addressed to authors and to reviewers, and we report what our case study offers against each in  \secref{sec:case:materials}.}
\Description{Table of four requirements for Time Machine Experiments: historical grounding, temporal integrity, interaction specification, and comparative attribution. Each row gives a guiding question, a way the inference can fail, and the evidence a study should report to discharge the requirement.}
\label{tab:validity}
\centering\small
\setlength{\tabcolsep}{5pt}
\begin{tabularx}{\linewidth}{@{}>{\raggedright\arraybackslash}p{0.2\linewidth}>{\raggedright\arraybackslash}X>{\raggedright\arraybackslash}X@{}}
\toprule
Requirement & How the inference fails & What to report \\
\midrule
\vreq{1} \textbf{Historical grounding.} What material grounds the system's representation of the period, and with what provenance? &
The model's output is read as what people of the period believed, when what it extends is a partial and unevenly preserved written record whose selection is inherited unstated~\cite{atari2023humans}. &
The corpus, its size and its boundary date; what the system is and is not claimed to represent; and any external check of the model's stances against contemporaneous evidence, and how long after the boundary that evidence dates from. \\
\addlinespace
\vreq{2} \textbf{Temporal integrity.} Does the boundary hold in the system as deployed, and not only in the base model? &
Post-boundary information enters through training data, prompts, retrieval, safety components, or other parts of the assembled system, and asking the model, or instructing it to stay in period, cannot establish that it has not~\cite{gao2025prompts}. &
Each component of the deployed system and its boundary status; which components generate text and which only filter it; whether the conditions differ in these components; and a probe of the assembled configuration, with its result. \\
\addlinespace
\vreq{3} \textbf{Interaction specification.} What did participants encounter, and how much of it was fixed before data collection? &
The model's voice is at once the stimulus participants predict and the evidence that the period differed, so choices about prompting, response selection, and item composition can manufacture the contrast the study then reports as a finding. Conditions may also differ in exposure in ways that were not intended. &
Stimulus provenance and any derivation from source material; the elicitation procedure and the alternatives tested against it; what is fixed before collection and what was generated live; and logs of unscripted interaction. \\
\addlinespace
\vreq{4} \textbf{Comparative attribution.} What does the comparison condition hold constant, and what can it not? &
An effect attributed to temporal boundedness is carried by differences that travel with it (e.g., style, verbosity, capability, refusal behavior) and cannot be separated from them, because no available model differs from the historically-bounded one in its cutoff alone. &
What is matched across conditions; the differences that are not matched, measured rather than asserted; and the scope of the causal claim, which is the bounded condition as a package, not the boundary in isolation. \\
\bottomrule
\end{tabularx}
\end{table*}

Stating the requirements also makes explicit the limits of what a Time Machine Experiment supports. Interaction with a historically-bounded model does not establish what people in the corresponding historical population believed, and an observed effect does not by itself isolate temporal boundedness from the properties that accompany it (\vreq{4}). Within those limits, the requirements offer a practical foundation for designing experiments and interactive systems around historically bounded encounters (see also \secref{sec:design}).

\section{Proof of Concept: Reversing the Illusion of Moral Decline}
\label{sec:case}

To demonstrate the potential of the Time Machine Experiment as an instrument for scientific inquiry, we conducted a proof-of-concept experiment targeting a measurable belief people today hold about the past.
Specifically, we selected the illusion of moral decline as a well-established phenomenon that has attracted attention across disciplines.
Mastroianni and Gilbert~\cite{mastroianni2023illusion} found that people consistently perceive people today as less kind, honest, and morally good than people in the past.
This finding comes despite evidence, based on repeated surveys (e.g., Gallup World Poll~\cite{tortora2010gallup}, General Social Survey~\cite{davern2025gss}, and World Value Survey~\cite{inglehart2022wvs}), that contemporaneous evaluations of morality have remained relatively stable over time, while experimental evidence suggests that cooperation among strangers in the United States has increased~\cite{yuan2022cooperation}.
Beyond the academic importance of the finding, such an illusion can, in principle, have practical consequences by directing public attention and resources toward reversing a decline that has not occurred.
It may also increase the appeal of political narratives promising to restore an idealized past (e.g., around women's reproductive rights).

The historically-bounded models allowed us to investigate and potentially alter this illusion experimentally through interaction with an interlocutor grounded in a pre-1930 corpus.
The model makes the encounter with historical perspectives more situated and responsive than an archive allows.
We examined whether such an encounter would update participants' perceptions of moral change relative to completing the same interaction with a contemporary frontier model, which served as a control condition.
This case study uses the Time Machine Experiment in both of its roles, as a method for studying how contemporary people respond to historically bounded perspectives and as an intervention for changing how they perceive the past.

\subsection{Design overview}

\begin{figure}
  \centering
  \includegraphics[width=\linewidth]{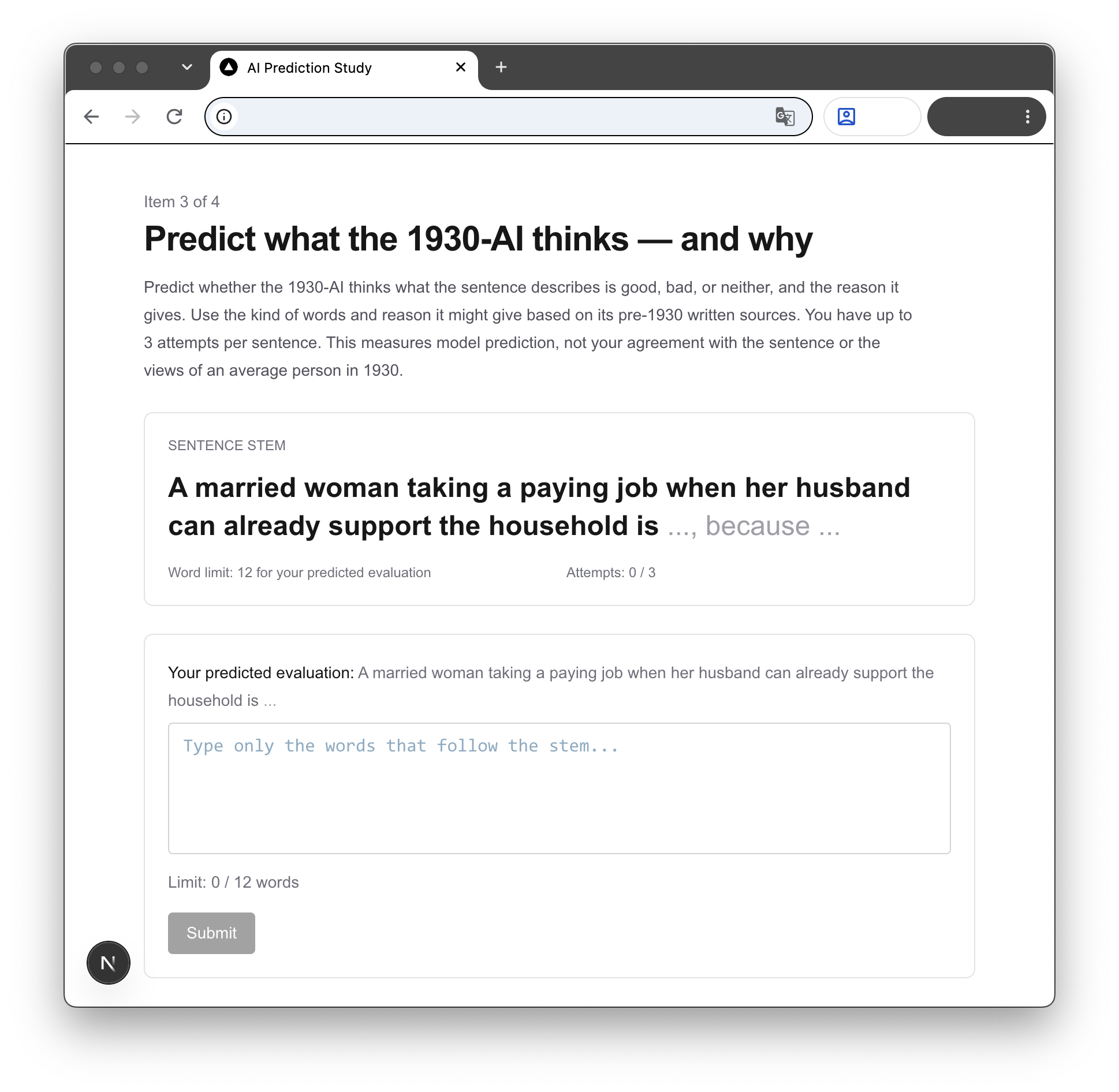}
  \caption{\textbf{Interface for the prediction.}
  Participants predicted whether their assigned model would evaluate the situation described by each sentence stem as good, bad, or neither, together with the reason it would provide.
  The task required participants to understand the model's perspective, and thus, they have an opportunity to ask arbitrary questions to the model beforehand.
  Participants could make up to three attempts for each task and received feedback after each attempt.}
  \Description{Screenshot of the prediction-task interface. Participants see a morally relevant sentence stem and enter their prediction of the assigned AI's evaluation and reasoning. The interface indicates the current item and attempt and provides feedback after submission.}
  \label{fig:prediction-interface}
\end{figure}

We conducted a preregistered randomized experiment comparing participants' perceptions before and after interactions with a historically-bounded or a contemporary AI model.
Participants completed a gamified prediction task in which they anticipated whether their assigned model would evaluate sentence stems describing morally and culturally contested situations as good, bad, or neither, as well as the reasoning it would provide (\figref{fig:prediction-interface}).
We used Talkie~\cite{talkie2026model}, a language model trained exclusively on a pre-1930 corpus, as the historical model, and \texttt{gpt-5.5}~\cite{openai2026introducing} as the contemporary control model, which we refer to as the \textit{1930-AI} and \textit{Modern-AI} conditions, respectively.
Both conditions used the same interface, sentence stems, task structure, and outcome measures, so that the conditions differed in the assigned model and in the responses it produced (\vreq{4}).
Participants were randomly assigned to either of the two conditions, which identifies the causal effect of the model on changes in participant evaluations.
The characteristics of the assigned model were disclosed only after the baseline responses had been recorded, so that condition framing could not influence pre-interaction judgments.

While participants had free-form conversations with their assigned model (see also \secref{sec:case-procedure}), we made the prediction task the core of the intervention for two reasons.
First, relying on unconstrained conversation alone would leave the content of the interaction largely uncontrolled.
Participants could spend the session discussing topics unrelated to moral judgment, making exposure difficult to standardize across participants.
The gamified prediction task instead provided a common set of morally relevant prompts while incentivizing participants to attend to and learn the model's perspective.
Second, the historically-bounded model has a smaller number of parameters compared to contemporary frontier models and is therefore less capable of maintaining coherence over long conversational contexts.
If extended free-form dialogue had been the primary intervention, differences in conversational coherence or context handling could have become an additional source of variation beyond the intended difference in temporal knowledge.
The structured prediction task therefore helped constrain the comparison to the model's evaluative perspective despite of broader differences in conversational capability.

In this experiment, participants completed the prediction task after an initial free-form conversation designed to familiarize them with their assigned model.
The prediction task consisted of four items randomly drawn from the predefined item pool.
For each item, participants could make up to three attempts and received feedback along with a performance score after each attempt, so that they could progressively build an understanding of the assigned model's perspective.
We measured perceived moral decline before and immediately after this interaction so that we could estimate how the pre-to-post change in perceived moral decline differed between conditions.

\subsection{Materials}
\label{sec:case:materials}

The 1930-AI is powered \texttt{talkie-1930-13b} \citep{talkie2026model}, which was pretrained on 260 billion tokens of English-language text published before 1930 (\vreq{1}).
The model was accessed through a moderated interface and presented to participants under the name 1930-AI.
We also compiled a list of items for the prediction task, which were to be evaluated by the AI and the participants, based on the following approach (see \appref{app:materials} for details).

The items used were fetched from a collection of closed-ended questions from two long-running survey programs that documented the history of public opinion: Gallup World Poll (GWP)~\cite{tortora2010gallup} first fielded between 1936 and 1969 and the General Social Survey (GSS)~\citep{davern2025gss}.
From an initial pool of approximately 420 candidate wordings we selected 12, requiring that each question (i) be intelligible to a pre-1930 context, (ii) ask for an evaluation on morally and not facts, (iii) carry no explicit moral vocabulary in the stem shown to participants, and (iv) be relevant in both eras of comparison (in this case, 1930 and today).

We therefore recast each question as a sentence to be completed, dropping the interrogative frame and leaving the sentence open mid-clause, and appended a second clause, \emph{because}, so that a completion must justify itself rather than assert a bare verdict. For instance, the 1936 wording ``Do you approve of a married woman earning money in business or industry if she has a husband capable of supporting her?'' becomes ``\stem{A married woman taking a paying job when her husband can already support the household is \underline{\ \ }, because \underline{\ \ }}'' (see \appref{app:reformulation} for other examples). We expected this structure to serve three purposes: to make the encounter engaging, since inferring an argument requires more cognitive effort than just guessing a verdict; to make positions comparable, since model and participant complete the identical object; and to facilitate scoring system described below, since stance and grounds occupy separate slots and can be weighed separately.

The prediction task asks participants to predict how their assigned model completes each item with a moral verdict and reasoning. Predictions are therefore made against fixed targets. For every item and model, we collected a single reference completion in advance, so that all participants in a condition are scored against the same response. Since models' probablistic behaviors do not necessarily hold one deterministic position on an item but a distribution over positions, we sampled repeatedly for each item and model, and selected completions that passed under basic quality checks: (i) coherent, in that the reason supports the verdict; (ii) not empty; (iii) not circular; (iv) understandable in present-day words; (v) concise in length; (vi) single-clause; and (vii) potentially guessable by a participant.
We then prioritized the position the sample most often took from the model (see \appref{app:modelresponseselection}).

Participants' predictions were scored against that reference on a 0--5 scale to facilitate their learning.
We used an LLM judge by applying a rubric calibrated against three human raters (Krippendorff's $\alpha = 0.87$), in which the stance fixes the band (i.e., opposite side, no side, or same side) and the closeness of the grounds moves the score within it (see \appref{app:scoring}).

\subsection{Procedure}
\label{sec:case-procedure}

\begin{figure*}
  \centering
  \includegraphics[width=\linewidth]{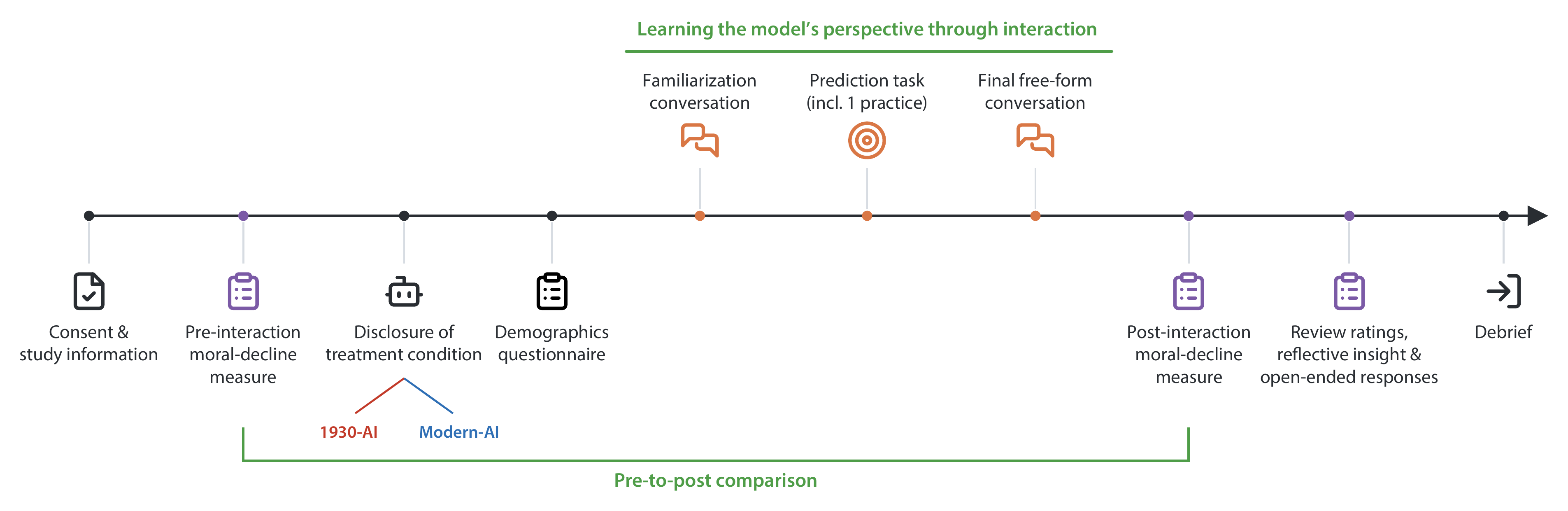}
  \caption{\textbf{Study procedure.}
  Participants completed the condition-neutral baseline measurement for perceived moral decline before receiving information about their assigned model.
  Their interaction with the model consisted of a brief familiarization phase, a practice item and four-item prediction task, and a final free-form conversation.}
  \Description{Horizontal timeline of the study procedure. Participants complete a baseline moral-decline measure before learning their assigned treatment condition, then interact through familiarization, a prediction task, and a final free-form conversation, followed by the post-interaction measure, reflection questions, and debriefing.}
  \label{fig:study-flow}
\end{figure*}

Figure~\ref{fig:study-flow} illustrates the study procedure.
Participants first completed the pre-interaction moral-decline measure (e.g., ``How much did people about 100 years ago value being kind, honest, nice, and good?'').
After that, they received information about their assigned model, followed by a comprehension check confirming their understanding of its training data.
The main component of the experiment was the prediction task described above, consisting of one practice item and four main items.
Before this task, participants familiarized themselves with their assigned model by conversing about two provided questions and one question of their own, allowing them to observe and probe its behavior.
Also, after the four prediction tasks, the participants engaged in a free-form conversation with their assigned model again, so that they could further probe its perspective after receiving feedback during the prediction task.
They subsequently completed the same moral-decline measure, as well as open-ended questions about their experience.
Finally, participants received a debriefing clarifying that the models' responses did not represent the views of the researchers and that the study did not seek to promote historical viewpoints.
The experimental system was implemented using Next.js, and its source code is publicly available at [anonymized link; included in the submission].

\subsection{Participants}

We recruited 240 English-speaking participants living in the United States or the United Kingdom through Prolific and randomly assigned them to either the 1930-AI ($N=121$) or Modern-AI ($N=119$) condition.
The sample size was preregistered on AsPredicted\footnote{\url{https://aspredicted.org/j99tb3.pdf}} based on a power analysis.
Prior to the main study, we conducted pilot sessions ($N=80$) to evaluate the technical functionality of the interaction and inform the sample-size determination for the main experiment.
Pilot data were not included in the analyses reported here.

The sample for the main study included 120 women (50.0\%), 118 men (49.2\%), one selected other (0.4\%), and one preferred not to report (0.4\%), and participants were aged from 19 to 83 years ($M=39.30$, $SD=12.62$).
Median session duration was 22 minutes, and they were compensated with \pounds3.80 for their participation, with an additional \pounds2 bonus for the top decile of prediction performance.
The study was approved by the institution's ethics review board committee.

\subsection{Measures}

To assess the changes targeted by the intervention, we measured perceived moral decline, reflective insight, prediction performance, and participants' open-ended responses.

\paragraph{Perceived moral decline.}
We measured perceived moral decline using items based on those used by Mastroianni and Gilbert~\cite{mastroianni2023illusion}.
Participants evaluated the moral values and behavior of people today, approximately 20 years ago, and approximately 100 years ago using seven-point scales.
Here, we extended the original measure, which asks how kind, honest, nice, and good people are, by deliberately distinguishing between value endorsement (\textit{How much do people today value being kind, honest, nice, and good?}; 1 = not at all, 7 = a great deal) and behavioral compliance (\textit{In their actual behavior, how well do people today live up to their own values about being kind, honest, nice, and good?}; 1 = not well at all, 7 = very well), each asked for today, about 20 years ago, and about 100 years ago.
We chose this distinction because a lower morality rating for the past could reflect either a belief that people placed less importance on being moral (our endorsement variable) or a belief that their behavior failed to live up to their own values (our compliance variable).

For each temporal horizon, we calculated a perceived-decline score by subtracting the rating of people in the past from the rating of people today, $D_{s,h} = \text{rating}_{s,\mathrm{today}} - \text{rating}_{s,\mathrm{past}(h)}$, for $s \in \{\mathrm{pre},\mathrm{post}\}$ and $h \in \{20,100\}$.
Negative values indicate perceived moral decline.
The change in perceived decline is $\Delta D_h = D_{\mathrm{post},h} - D_{\mathrm{pre},h}$, so that positive values indicate a weaker illusion after the conversation.
Our primary outcome was $\Delta D_{100}$ for endorsement, compared between conditions with a Welch $t$-test. For example, a participant who answers 4 to ``How much do people today value being kind, honest, nice, and good?'' and 6 to the same question about people 100 years ago has $D_{\mathrm{pre},100} = 4 - 6 = -2$; if the person answers 4 and 5 after the conversation, $D_{\mathrm{post},100} = -1$ and $\Delta D_{100} = +1$, meaning the gap the person perceives between past and present narrowed by one scale point.
We then regressed the same outcome on condition and the demographic variables collected before the interaction (\tabref{tab:ols}).
As a complementary direct measure of perceived change, participants also directly indicated whether endorsement and compliance were greater, lower, or approximately the same today as 100 years ago.

\paragraph{Reflective insight.}
We measured reflective insight using three items adapted from the Insight subscale of the Technology-Supported Reflection Inventory (TSRI)~\cite{bentvelzen2021tsri}.
We used this subscale in TSRI because it directly captures whether an interactive technology prompts new perspectives or changes in thinking. 
Participants indicated whether the interaction made them reconsider their own perspectives and values, changed how they viewed the past and present, and allowed them to see the world from a perspective they had not previously considered.
Responses were provided on a seven-point scale from 1 (\textit{strongly disagree}) to 7 (\textit{strongly agree}).

\paragraph{Prediction performance.}
Participants predicted how their assigned AI would evaluate four morally or culturally relevant sentence stems, including both the evaluation itself and the reason the AI would provide.
As above, they could make up to three attempts per item and received feedback after each attempt.
Predictions were scored from 0 to 5 by the LLM judge, and for each item, we retained the participant’s highest-scoring attempt and calculated prediction performance as the mean of these best-attempt scores.
This measure determined eligibility for the performance bonus but was not a primary outcome of the intervention analysis (see \appref{app:prediction-similarity}).

\paragraph{Open-ended responses.}
We also analyzed the content that participants asked their assigned AI during a free-form conversation at the final stage and subsequently responded three open-ended questions concerning what surprised them about the AI’s evaluations, whether the interaction changed how they thought about their values or moral change over time, and any additional feedback. We used these qualitative data to contextualize the quantitative findings and analyzed these data descriptively rather than through a formal coding scheme.
One author read all 240 sets of open-ended responses and all final-conversation transcripts in full, grouped recurring content inductively by topic, and noted for each group how many participants in each condition it appeared in.
Quotes reported below were chosen as typical of their group rather than as the most striking instances.
No inter-rater reliability was computed, and the qualitative material serves to contextualize the quantitative results rather than to support claims of its own.

\subsection{Results}

\subsubsection{Effects on perceived moral decline}

\begin{figure*}
  \centering
  \includegraphics[width=\linewidth]{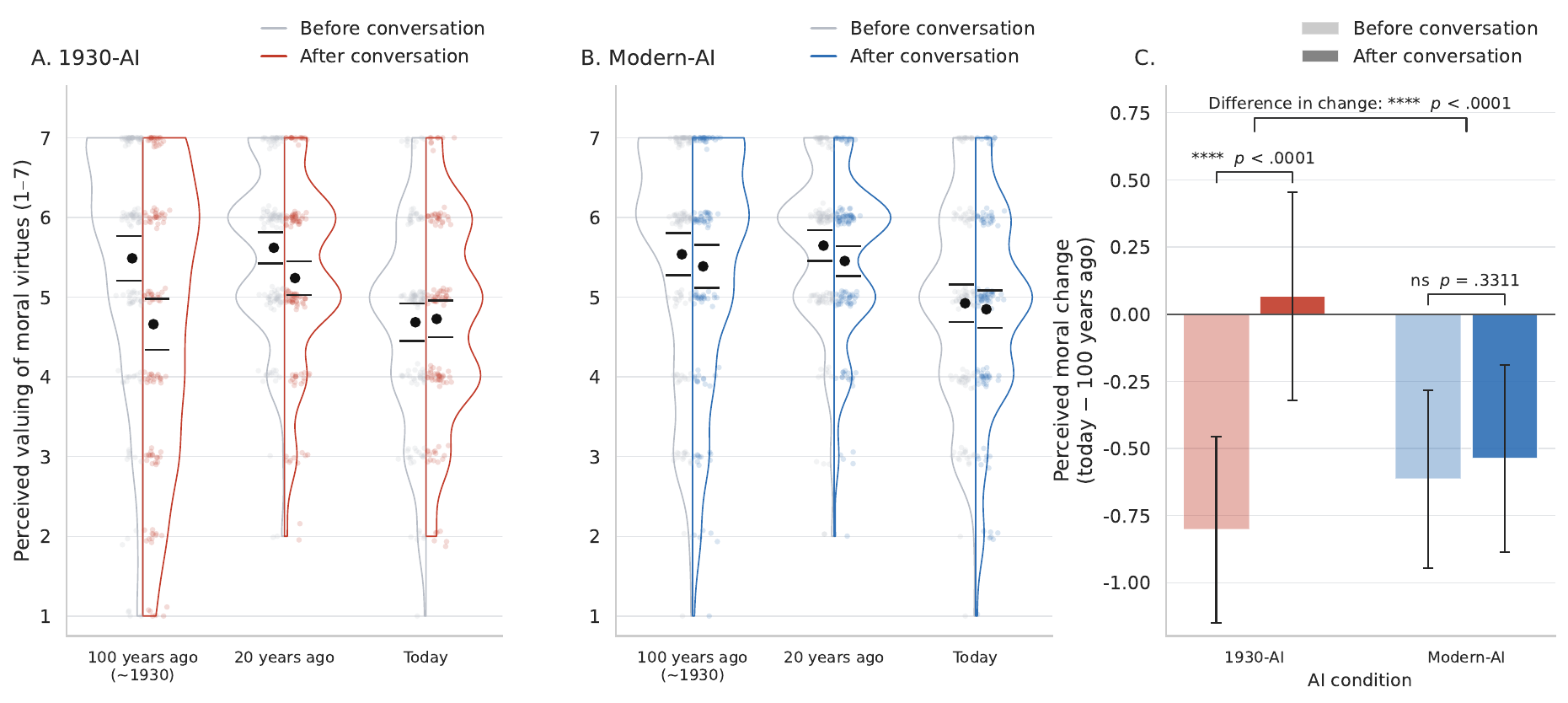}
  \caption{\textbf{Moral-value endorsement before and after the AI conversation.}
  Panels \textbf{(A)} and \textbf{(B)} show how much participants believed people valued being kind, honest, nice, and good at three time horizons, in the 1930-AI and Modern-AI conditions, respectively.
  Split violins show response distributions; black points and lines indicate means and 95\% confidence intervals.
  Panel \textbf{(C)} shows perceived moral decline, the today rating minus the 100-year rating, so negative values indicate perceived decline, before and after the conversation by condition.
  Short brackets report paired-samples $t$-tests comparing ratings before and after the conversation within each condition; the upper bracket reports the between-condition difference in participant-level pre-to-post change, tested with Welch's independent-samples $t$-test. Significance levels are $^{*}p<.05$, $^{**}p<.01$, $^{***}p<.001$, and $^{****}p<.0001$.}
  \Description{Three-panel result figure for moral-value endorsement. (A) and (B) are split violin plots of endorsement ratings at three time horizons, before and after the conversation, for the 1930-AI and Modern-AI conditions. (C) shows before and after perceived-decline scores by condition as bars, with short brackets marking within-condition tests and an upper bracket marking the between-condition test.}
  \label{fig:main-results}
\end{figure*}

Before the conversation, participants in both conditions perceived moral decline over the preceding 100 years (\figref{fig:main-results}).
Ratings of how much people valued being kind, honest, nice, and good were higher for people 100 years ago than for people today, in the 1930-AI condition (\figref{fig:main-results}A) and in the Modern-AI condition (\figref{fig:main-results}B), and the resulting decline scores were negative in both (1930-AI: $M=-0.80$, 95\% CI $\left[-1.15, -0.45\right]$; Modern-AI: $M=-0.61$, 95\% CI $\left[-0.94, -0.28\right]$; \figref{fig:main-results}C).
The conditions did not differ at baseline (Welch $t(237.65)=-0.78$, $p=.438$).\footnote{$M$ and $SD$ denote the sample mean and standard deviation, $CI$ a 95\% confidence
interval, $t$ a $t$ statistic with degrees of freedom in parentheses (fractional where Welch's
correction for unequal variances is applied), $p$ a two-sided $p$ value and $p_{\mathrm{adj}}$ its
Holm-adjusted counterpart, and $d$ Cohen's $d$ for between-condition differences with $d_z$ its
within-participant counterpart.}

Descriptively, the conversation moved the ratings of the past.
In the 1930-AI condition, the rating of people 100 years ago fell from 5.49 to 4.66 (paired $t(120)=-6.21$, $p<.001$, $d_z=-0.56$), the rating of people 20 years ago fell less, from 5.62 to 5.24 ($p<.001$), and the rating of people today did not change ($p=.631$; \figref{fig:main-results}A).
In the Modern-AI condition, ratings of people 100 years ago fell from 5.54 to 5.39 ($p=.003$) and ratings of people 20 years ago fell from 5.65 to 5.45 ($p<.001$), whereas ratings of people today did not change ($p=.226$; \figref{fig:main-results}B).
The largest distributional shift was the downward spread of responses at the 100-year horizon after the conversation with the 1930-AI.

The change in perceived decline, $\Delta D_{100}$, was larger in the 1930-AI condition ($M=0.87$, $SD=1.80$) than in the Modern-AI condition ($M=0.08$, $SD=0.85$); the difference was 0.79 points (95\% CI $[0.44, 1.15]$, Welch $t(171.21)=4.38$, $p<.001$, $d=0.56$), marked by the upper bracket in \figref{fig:main-results}C.
Within conditions, the decline score rose from $-0.80$ to $0.07$ in the 1930-AI condition (paired $t(120)=5.31$, $p<.001$, $d_z=0.48$) and from $-0.61$ to $-0.54$ in the Modern-AI condition (paired $t(118)=0.98$, $p=.331$, $d_z=0.09$).
In the 1930-AI condition, the small positive post-conversation mean indicates that the perceived decline disappeared and slightly reversed. Because ratings of people today did not change significantly, the change came from lower ratings of the past rather than higher ratings of the present.

The pattern held for moral compliance, the second measure (Appendix \figref{fig:moral-trajectories}).
The change was larger in the 1930-AI condition ($M=0.82$, $SD=2.01$) than in the Modern-AI condition ($M=0.21$, $SD=0.91$; difference $0.61$, 95\% CI $[0.21, 1.01]$, Welch $t(167.73)=3.02$, $p=.003$, $d=0.39$), and the decline score rose from $-1.05$ to $-0.23$ in the 1930-AI condition ($p<.001$) and from $-0.86$ to $-0.65$ in the Modern-AI condition ($p=.013$).
Both between-condition differences survived Holm correction across the four decline outcomes ($p_{\mathrm{adj}}<.001$ for endorsement and $p_{\mathrm{adj}}=.009$ for compliance).

The between-condition differences were larger at the 100-year horizon but were also detectable at the 20-year horizon for endorsement (difference $0.30$, 95\% CI $[0.03, 0.58]$, $p_{\mathrm{adj}}=.036$) and compliance (difference $0.30$, 95\% CI $[0.05, 0.54]$, $p_{\mathrm{adj}}=.036$).
Participants' direct comparative judgments corroborated this: the proportion judging morality as lower today than 100 years ago fell more in the 1930-AI than in the Modern-AI condition for endorsement (61.2\% $\rightarrow$ 40.5\% vs.\ 53.8\% $\rightarrow$ 50.4\%) and compliance (57.9\% $\rightarrow$ 41.3\% vs.\ 58.8\% $\rightarrow$ 54.6\%).

\subsubsection{Robustness to participant characteristics}

To check that the treatment difference does not depend on who happened to be assigned to which condition, we fit an ordinary least squares regression with the change in perceived moral decline as the outcome, that is, each participant's perceived moral decline (the difference between the today and the 100-years-ago rating) after the conversation minus the same score before it, and condition as the predictor of interest, adding the background variables collected before the interaction as controls:
\begin{equation}
\Delta D_{100,i} = \beta_0 + \beta_1\,\mathrm{Condition}_i + \boldsymbol{\beta}_{\mathrm{c}}^{\top}\mathbf{X}_i + \varepsilon_i,
\label{eq:ols}
\end{equation}
where $\mathrm{Condition}_i$ is 1 for the 1930-AI and 0 for the Modern-AI, and $\mathbf{X}_i$ collects age, gender, education, country of residence, political self-placement, and AI chatbot use.
The coefficient $\beta_1$ is the difference in change between conditions after adjustment; without the controls it equals the Welch $t$-test difference reported above.
Under random assignment the background variables should be balanced across conditions, so their inclusion should change $\beta_1$ little; a large shift would indicate chance imbalance.
We fit the same specification separately for endorsement and compliance (\tabref{tab:ols}), with heteroskedasticity-robust (HC3) standard errors.
The adjusted condition effect remained positive for endorsement ($b=0.80$, HC3 $SE=0.18$, 95\% CI $[0.44, 1.16]$, $p<.001$) and compliance ($b=0.65$, HC3 $SE=0.20$, 95\% CI $[0.26, 1.05]$, $p=.001$), close to the unadjusted differences.
Identifying as a woman ($b=0.46$, $SE=0.20$, $p=.019$) and residing in the UK ($b=0.64$, $SE=0.21$, $p=.002$) were associated with larger changes in perceived compliance decline; no other covariate was significant in either model.

\begin{table}[t]
\caption{OLS regressions of the change in perceived 100-year moral decline ($\Delta D_{100}$; positive values indicate a weaker illusion after the conversation) on condition and participant characteristics, Eq.~\ref{eq:ols}. Condition is 1 for 1930-AI and 0 for Modern-AI, so a positive coefficient means a larger reduction of the illusion in the 1930-AI condition; the decline score rises toward zero as the illusion weakens. Woman is 1 for participants who reported their gender as female and 0 otherwise. Education is the six-level self-report from 1 (did not finish high school) to 6 (graduate degree). UK is 1 for participants residing in the United Kingdom and 0 for the United States. Ideology is self-placement from 1 (left-wing) to 7 (right-wing). AI use is frequency of AI chatbot use from 1 (never) to 5 (daily). One impossible age response (1851) was treated as missing, so both models use 239 complete cases. Heteroskedasticity-robust (HC3) standard errors are in parentheses. $^{*}p<.05$, $^{**}p<.01$, $^{***}p<.001$.}
\Description{Regression table with separate models for moral endorsement and compliance. Each model predicts the pre-to-post change in perceived 100-year moral decline from AI condition and participant characteristics. The positive 1930-AI condition effect remains significant after adjustment in both models.}
\label{tab:ols}
\centering\small
\begin{tabular}{@{}lcc@{}}
\toprule
 & Endorsement & Compliance \\
\midrule
Condition (1930-AI) & 0.80 (0.18)$^{***}$ & 0.65 (0.20)$^{**}$ \\
Age & 0.00 (0.01) & $-0.01$ (0.01) \\
Woman & 0.26 (0.18) & 0.46 (0.20)$^{*}$ \\
Education & $-0.01$ (0.06) & $-0.01$ (0.07) \\
UK & 0.31 (0.20) & 0.64 (0.21)$^{**}$ \\
Ideology & $-0.06$ (0.05) & $-0.09$ (0.06) \\
AI use & 0.09 (0.11) & 0.18 (0.11) \\
Intercept & $-0.51$ (0.62) & $-0.42$ (0.76) \\
\midrule
$N$ & 239 & 239 \\
$R^2$ & 0.10 & 0.12 \\
\bottomrule
\end{tabular}
\end{table}

\subsubsection{Reflective insight}

\begin{figure}
  \centering
  \includegraphics[width=\linewidth]{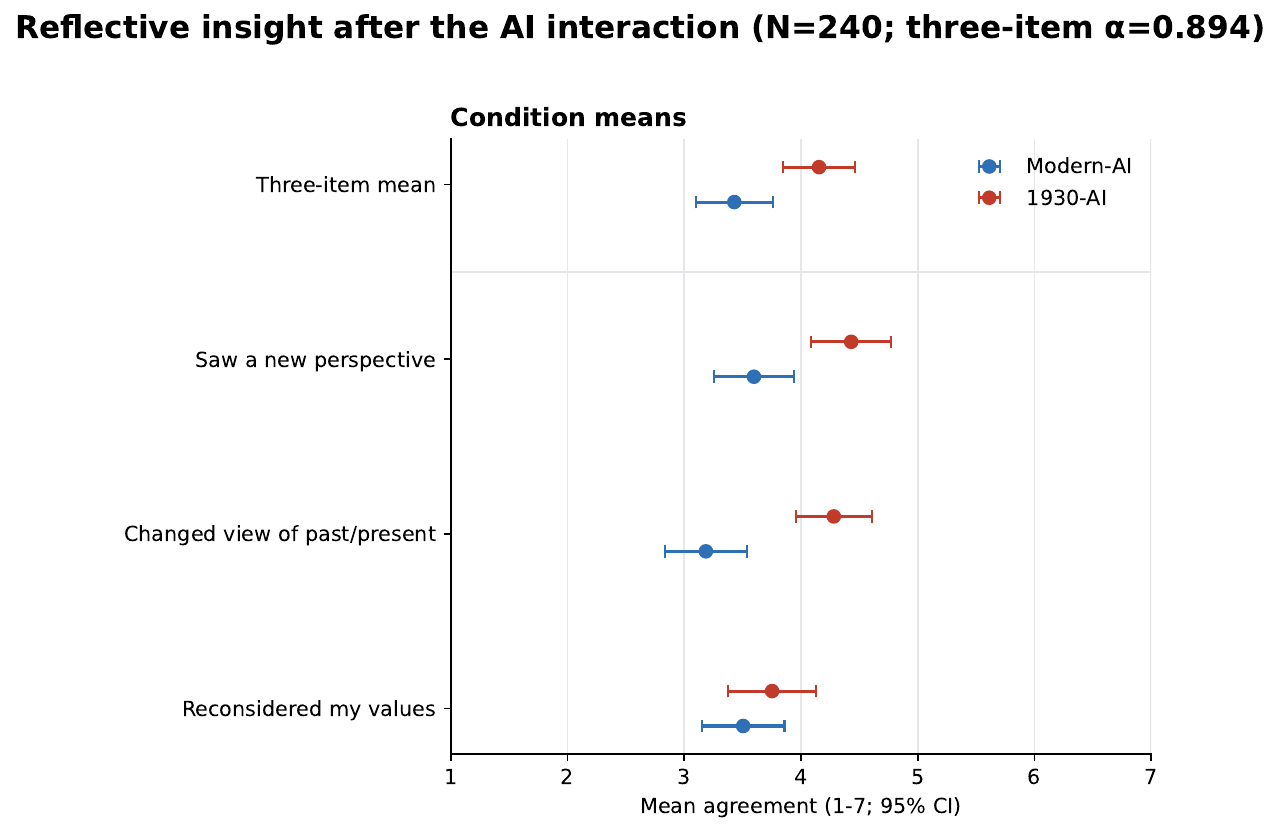}
  \caption{\textbf{Reflective insight after the AI interaction.}
  Mean agreement with the three reflective-insight items and their composite, shown separately for the Modern-AI and 1930-AI conditions.
  Responses ranged from 1 (\textit{strongly disagree}) to 7 (\textit{strongly agree}).
  The horizontal line separates the three-item mean from its constituent items.
  Points indicate condition means and error bars show 95\% confidence intervals.}
  \Description{Dot-and-error-bar plot comparing reflective insight between conditions. It has four rows: the three-item mean above a divider, then the three individual insight items below it. Each row shows a red point for the 1930-AI condition and a blue point for the Modern-AI condition with horizontal 95 percent confidence intervals on a 1 to 7 scale. The 1930-AI points sit to the right of the Modern-AI points on all rows, with the largest gap for the changed view of past and present item and the smallest for reconsidered my values.}
  \label{fig:insight-means}
\end{figure}

The three reflective-insight items were internally consistent ($\alpha=.894$) and were averaged into a composite.
Participants in the 1930-AI condition reported greater reflective insight ($M=4.15$, $SD=1.70$) than those in the Modern-AI condition ($M=3.43$, $SD=1.82$; difference $0.73$, 95\% CI $[0.28,1.17]$, Welch $t(238)=3.19$, $p=.002$, $d=0.41$; \figref{fig:insight-means}).
At the item level, the 1930-AI condition produced higher ratings for changing how participants viewed the past and present ($\Delta M=1.10$, $p_{\mathrm{adj}}<.001$, $d=0.59$) and for seeing the world from a previously unconsidered perspective ($\Delta M=0.83$, $p_{\mathrm{adj}}=.002$, $d=0.44$), but not for reconsidering one's own perspectives and values ($\Delta M=0.25$, $p_{\mathrm{adj}}=.346$, $d=0.12$).

Taken together with the above results, interacting with the 1930-AI reduced the illusion of moral decline over the 100-year horizon across the population and elicited greater reflective insight than interacting with the Modern-AI.
The absence of comparable effects for the 20-year outcomes is consistent with the intended temporal focus of the 1930-AI, whose training corpus exposed participants to perspectives from approximately a century earlier. This indicates that the interaction with historically-bounded models can measurably shape human perception and reflection, providing an initial proof of concept for the Time Machine Experiment as a method for both inquiry and intervention across temporal contexts.

\subsubsection{Performance in predicting the AI's responses}

Participants predicted the evaluations of the Modern-AI more accurately ($M=4.43$, $SD=0.42$) than those of the 1930-AI ($M=4.27$, $SD=0.55$).
The mean difference was 0.17 points (95\% CI $[0.04, 0.29]$, $t(224.26)=2.66$, $p=.008$, $d=0.34$), indicating that the historically trained model was relatively more difficult to predict, which aligns with what we would expect.
Alternative embedding- and likelihood-based similarity measures, together with examples of participants' predictions, are reported in Appendix~\ref{app:prediction-similarity}.
On the other hand, prediction performance was not associated with reflective insight ($r=-.03$, $p=.646$) or with change in perceived moral decline ($r=.08$, $p=.225$).
This indicates the absence of associations between prediction performance and either reflective insight or belief updating, which suggests that the reflective effects of the 1930-AI arose less from successfully learning to predict the model and more from encountering unfamiliar historical perspectives.
More broadly, this finding again underlines the prospective value of the Time Machine Experiment to encourage participant engagement by exposing participants to perspectives that are neither familiar nor trivially predictable.
Such encounters may be particularly valuable for this line of research, as participants are increasingly exposed to repetitive and abstract online surveys.

\subsubsection{Qualitative responses}

The qualitative responses suggest that the 1930-AI prompted reflection more than it changed values.
Participants frequently reported that their core values remained unchanged, while describing a greater awareness of how moral judgments depend on historical and social context (e.g., ``It showed me morality is relative to the time you live in. It made me wonder what we believe now that future generations will find shocking.'').
The contrast with the historical model also prompted participants to reconsider moral progress, particularly concerning gender, race, and sexuality (e.g., ``It just made me think about how little freedom people had and how hard it was to be a woman,'' and ``The world isn't great today but it has drastically improved.'').
By comparison, participants described the Modern-AI as familiar and predictable and as broadly aligned with their existing values, although some read its emphasis on autonomy, equality, and harm prevention as a liberal bias (e.g., describing it as having a ``strong liberal bias and a lack of material consideration of other views.'').
Consistent with this, participants in the 1930-AI condition held longer final conversations and wrote longer reflections ($M=2.45$ vs.\ $1.66$ user turns; $M=27.6$ vs.\ $19.8$ words in their queries).

The questions participants put to the models during the free-form conversation show what they wanted to know about their assigned AI.
Participants in the 1930-AI condition frequently tested the limits of the model's historical worldview, asking about women's rights and social roles, racial equality and interracial relationships, homosexuality and same-sex marriage, abortion, religion, and immigration (e.g., ``Are women considered equal to men in society?'', ``Do you believe that every race is equal?'', and ``Is homosexuality permitted?'').
Participants often followed up by challenging the model's reasoning or pointing out inconsistencies between its abstract endorsement of equality and its judgments about particular social groups.

In contrast, participants in the Modern-AI condition more often asked about the general principles behind its judgments, such as harm, fairness, autonomy, honesty, and individual rights.
They also asked for its position on contemporary contested issues, including abortion, the death penalty, immigration, transgender rights, climate change, and electoral politics.
Several explicitly examined the model's political orientation or alignment with their own values, asking questions such as, ``What moral framework or ethical values guide your responses?'', ``How do your moral judgments compare with the average person's beliefs?'', and ``What political way do you lean?''
The contrast shows what the historically-bounded model adds.
Participants used it to question past norms and compare them with present principles, and in doing so reconsidered how far morality depends on time and place.

\section{Application space of Time Machine Experiments}
\label{sec:design}

\begin{figure*}[t]
  \centering
  \includegraphics[width=0.7\linewidth]{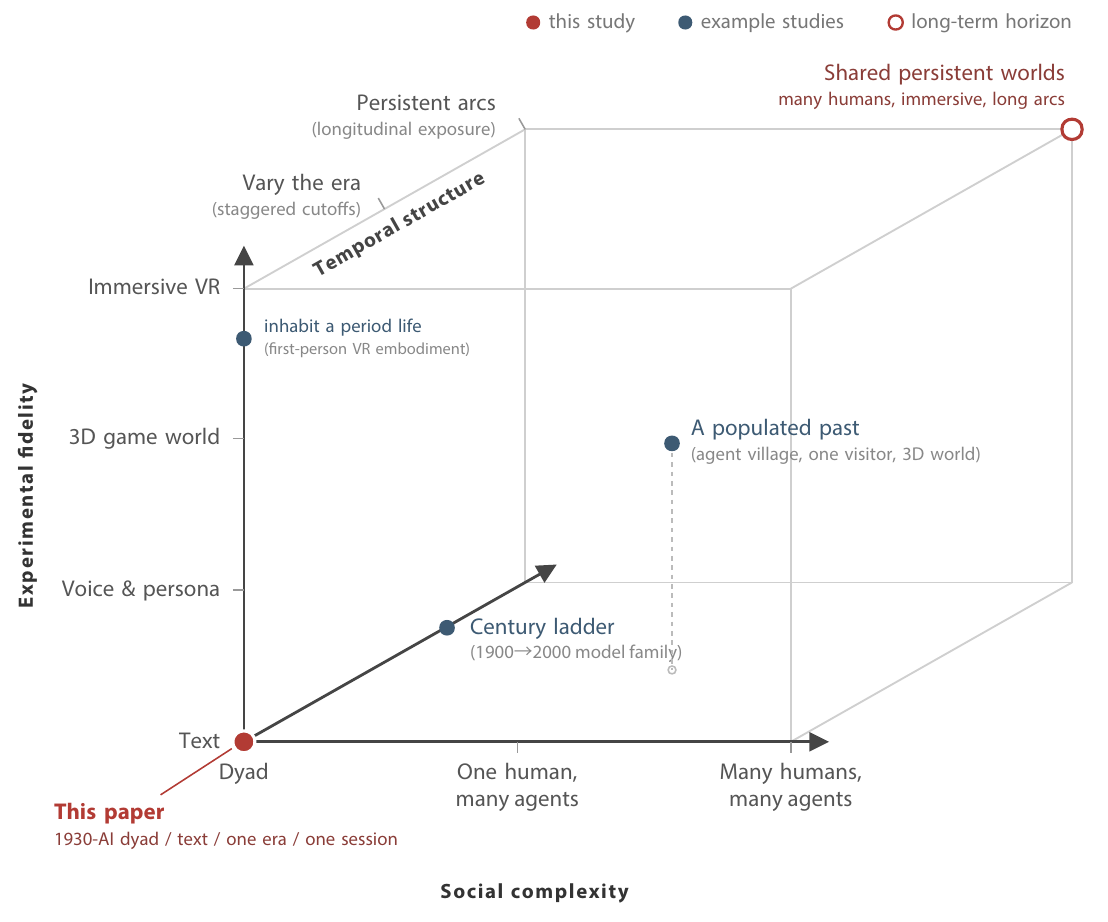}
  \caption{\textbf{The design space of Time Machine Experiments.}
  It spans social complexity, temporal structure, and experiential fidelity.
  Each experiment is a coordinate in this space.
  The present study occupies the origin, a dyadic, single-session, text interaction with a model of one era, where the validity requirements (\vreq{1}--\vreq{4}) are most tractable.
  Example studies mark points further out, and the open marker denotes a long-term possibility.}
  \Description{A three-dimensional axis diagram. The horizontal axis is social complexity from dyad to one human with many agents to many humans with many agents. The vertical axis is experiential fidelity from text through voice and persona and 3D game world to immersive VR. The depth axis is temporal structure from one era through varied eras to persistent arcs. A red point at the origin marks this paper. Blue points mark three example studies: a century ladder of models along the temporal axis, a populated past at higher social complexity and fidelity, and inhabiting a period life at the top of the fidelity axis. An open red circle at the far corner marks shared persistent worlds as a long-term horizon.}
  \label{fig:design-space}
\end{figure*}

While the proof of concept demonstrated the potential of Time Machine Experiments as research instruments, it occupies one point in a larger design space with three dimensions (\figref{fig:design-space}).
Social complexity runs from the dyad studied here, through one human among many era-bounded agents, to many humans and many agents, as in the populated past of the figure, where period norms are enforced among the agents through observation, gossip, and sanction rather than stated by a single interlocutor.
Temporal structure has two parts.
The training boundary can be moved, so that a family of models with staggered cutoffs forms a century ladder from 1900 to 2000, and the participant's exposure can run from a single session to a persistent arc of repeated contact.
Experiential fidelity rises from text, as here, through voice and persona and navigable game worlds to immersive VR in which a participant inhabits a period life in the first person.
Each step outward makes a validity requirement harder to establish.
Populations of agents produce interaction that cannot be fixed in advance (\vreq{3}), repeated sessions give the present more chances to leak in (\vreq{2}), earlier or sparser cutoffs weaken the record a model can be checked against (\vreq{1}), and a ladder of staggered cutoffs is the one move that makes the boundary separable from the other properties of a model (\vreq{4}).
This is why the origin is where the proof of concept sits, namely that the origin is where all three are most tractable.

Read as science fiction science, a Time Machine Experiment asks what contact with a standpoint from another time would do to a person, and runs the experiment before any technology for such contact exists.
Read plainly, the tools are already here.
A game designer can ground a character in a period corpus rather than in a writer's impression of one, and an educator can put a period model in front of a class; the experiment measures what those encounters do.
We can point three immediate applications.
In education and heritage, learners question a period model in classrooms, museums, or archives, and history becomes something that can be asked rather than only read.
Because the model's divergences from the record are documented, the same exercise teaches historical literacy and AI literacy.
As an intervention, contact with the past can act on beliefs beyond moral decline.
It can weaken golden-age nostalgia, and it may work generatively, since unfamiliar assumptions provoke new lines of thought and some present problems had workable answers that dropped out of collective memory and can be recovered by asking.
In entertainment, an era-bounded model supports games in which the player's goal is to change the past and the only means is to understand it.
Persuasion succeeds only through arguments intelligible in period terms, so progress through the game is itself a behavioral measure of perspective-taking.
What these settings share is a standpoint fixed at a date and made available for interaction.

\section{Discussion}
\label{sec:discussion}

\subsection{How did the time machine affect participants?}

In the 1930-AI condition, participants rated people 100 years ago less positively after the interaction, while their ratings of people today did not change (\figref{fig:main-results}A). The change therefore appears to have come primarily from a revised view of the past rather than a more positive view of the present. Two other results help clarify this pattern. The between-condition difference was largest at the 100-year horizon and smaller, though still detectable, at 20 years, a period about which the corpus contains no information. The effect was thus strongest at the historical period represented by the model. In addition, endorsement changed more than compliance: participants changed their judgments about what people in 1930 valued more than their judgments about how closely people at the time lived according to those values.

This pattern is broadly consistent with the account of the illusion proposed by Mastroianni and Gilbert~\cite{mastroianni2023illusion}. In their account, the belief that people were better in the past can arise from biased exposure to negative information about contemporaries and from a form of memory in which the negative aspects of the past fade faster than the past itself. Neither mechanism maps directly onto the present setting. Participants did not have memories of 1930, and their impressions of the period necessarily came from information available to them. The corpus therefore provided a new source of information about the period, but it was a particular kind of source, as it consisted of historical texts and primarily captured what people wrote and publicly expressed, rather than how they behaved. This distinction is relevant given that endorsement, and not compliance, showed the larger change. The gradient across horizons fits the same account. The model had no information about 2006, so the conversation supplied no new source for that period; the smaller 20-year effect can instead arise from the revised 100-year anchor, since a participant who lowers the 1930 endpoint has reason to adjust the point that lies between it and today.

Participants' own reports provide a similar indication of what changed. Those in the 1930-AI condition were more likely to agree that the conversation had changed how they saw the past and present ($d=0.59$) and that it had shown them a perspective they had not previously considered ($d=0.44$). There was little evidence that it made them reconsider their own values ($d=0.12$). The open-ended responses point in the same direction. Participants generally did not describe changing their values; instead, they described becoming more aware that those values belonged to a particular historical context. This distinction matters for interpreting the main result. The interaction appears to have shifted the historical comparison against which participants viewed their own values, without substantially changing those values themselves.

Prediction accuracy provides a further qualification. Participants predicted the 1930-AI less accurately than the Modern-AI, but prediction accuracy was not associated with either belief change or reported insight. This makes it less likely that the observed effects depended on participants becoming better at modeling the system. The null associations cannot establish the mechanism, however. They are simply consistent with the possibility that exposure to the period's expressed positions, rather than successful prediction of the model, contributed to the change.

\subsection{The boundary as an interaction variable}

The questions participants asked also differed between conditions. Those who interacted with the 1930-AI often asked about the rights and standing of particular groups, including women, racial minorities, sexual minorities, and immigrants. Participants in the Modern-AI condition were more likely to ask about the principles underlying the model's judgments or about its politics.

This difference suggests that the temporal boundary affected not only the answers participants received but also what they chose to investigate. In the 1930-AI condition, participants appeared to probe the model at points where they expected historical attitudes to diverge from contemporary ones. At the same time, this observation limits how broadly we should interpret the change in the general belief measure. The domains participants selected are precisely those in which differences between past and present are particularly visible. It is therefore possible that the change in beliefs about kindness and honesty was partly driven by exposure to these specific domains rather than by the historical standpoint more generally.

Prototypes~\cite{buchenau2000experience,boer2012provotypes}, speculative enactments~\cite{elsden2017on,soro2019designing,kozubaev2020expanding}, and immersive environments~\cite{blascovich2002immersive,miller2021synchrony,pi2025embodied} have long been used to make otherwise inaccessible situations available for experience. In these approaches, however, the situation that participants encounter is generally specified in advance. A conversational model changes this relationship because participants can ask questions that were not anticipated by the designer. HCI research has similarly used conversational systems to explore imagined future voices~\cite{lee2021conversational}. A historically bounded model extends this approach to the past, while introducing one difference. Some of its responses can be compared with surviving historical evidence. The interaction is therefore not purely evocative; at least some of what the participant encounters can be checked against the historical record.

This also suggests a broader role for knowledge boundaries in interaction design. A temporal boundary does more than constrain the model's knowledge. It shapes what participants can ask, what they expect the system to know, and where they notice gaps between the model's standpoint and their own. This is similar in spirit to treating machine-learning uncertainty as a design material~\cite{benjamin2021machine}. In the present study, the absence of later knowledge created part of the experimental condition, and participants could actively explore that absence through conversation. The same idea could extend beyond historical time. A model could be bounded by a particular culture, discipline, institution, or region, allowing the standpoint itself to become part of what an interaction assigns to the participant.

\subsection{Limitations}

A historically bounded language model should not be treated as a person sampled from a historical population. One reason is the nature of the historical record on which such a model depends. The ability to write, publish, and preserve texts was unevenly distributed, meaning that surviving corpora over-represent literate and institutionally connected groups while providing less evidence about everyday speech and marginalized populations. These biases have long been documented in historical scholarship~\cite{trouillot1995silencing}, and they can be compounded through archival survival, digitization, OCR, corpus construction, and model training. The resulting system is therefore better understood as a generative extension of what was written and preserved than as a representation of what ``people in 1930'' believed. For the same reason, the present experiment should not be interpreted as estimating what would happen if a contemporary participant actually met someone from 1930. The causal comparison is between a historically bounded model and a contemporary model; ``meeting the past'' is a motivating metaphor, not a description of the treatment (\vreq{1}).

The models also differed in ways beyond their temporal scope. These included size and training regime, linguistic style, capability, predictability, and what participants knew about the system. Participants in the 1930-AI condition learned from the model description and the comprehension check that they were interacting with a historical model. The Modern-AI condition therefore provides a control for interacting with an AI system, but not for knowing that the system is temporally bounded. The design consequently does not isolate temporal boundedness from the other features of the interaction. It is more appropriate to interpret the study as estimating the effect of the bounded interaction condition as a package. A closer comparison is possible with the Talkie release, which includes a model with the same architecture and training budget trained on contemporary web text~\cite{talkie2026model}. Such a comparison could hold size, style, and capability more constant while varying the temporal boundary. Staggered temporal cutoffs and style-normalized stimuli could further help separate these factors.

Finally, the effect was measured only minutes after a single interaction. We do not know whether the change persists over time or after repeated exposure. A follow-up study should test whether the effect remains at a delayed measurement and whether repeated encounters strengthen or diminish it. More broadly, this proof of concept examines one temporal boundary, one English-language model, participants from two countries, and one class of beliefs about the past. Whether the same interaction produces comparable effects across other historical periods, languages, domains, and beliefs remains an open question.

\subsection{Future Directions}

A stronger test of the present result would compare the 1930-bounded model with a matched contemporary model, measure whether the effect persists after a delay, and include the trust and honesty items on which the original finding rests, and each of these follows directly from the limitations above.
The instrument itself can also be extended in two ways, one that narrows the boundary and one that multiplies it.

A model continued on one person's writings up to a specified date, with that person's later writings held out of training, would narrow the boundary from an era to an individual, and the same could be done for a community or an institution.
The held-out writings would then serve as an independent test of what the model learned, which gives the historical-grounding requirement (\vreq{1}) a direct empirical form; an era-level model, by contrast, can be checked only against aggregate sources such as period polls, and only for the items where such a poll exists.

Starting from the same 1930-bounded model, two versions could instead be continued toward 1950 on different records, one including material about the Second World War and one trained on a corpus from which the war had been removed, which would have to be constructed rather than filtered since nearly all text from the period refers to the war.
Participants could then question the two models and compare how their responses diverge from a shared starting point.
Neither model would represent a possible history, and the design does not claim that; what it provides is two conversational counterparts that differ only in their subsequent information, which could be used to study how people respond to divergent trajectories and whether such encounters change their judgments about what was contingent and what was inevitable, extending speculative enactment and science fiction science~\cite{rahwan2025science} from scenarios authored by the researcher to scenarios generated from a common historical base.
The cost is a validity problem the present study did not face, because the counterfactual branch has no historical record against which its responses can be checked and so cannot satisfy \vreq{1}, leaving the comparison to rest on the boundary holding (\vreq{2}) and on the interaction being fully specified (\vreq{3}), which makes those two requirements more demanding than they were here.

\section{Conclusion}

We introduced \textit{Time Machine Experiment}, a methodological framework that uses historically-bounded language models as interactive instruments for studying and intervening in human perception, reasoning, and behavior across temporal contexts.
As a proof of concept, we conducted a preregistered randomized experiment showing that interaction with a model trained only on pre-1930 text reduced the illusion of moral decline and elicited greater reflective insight than interaction with a contemporary model.
These findings show that historically bounded encounters can be built, validated, and used to change how people see the past, and they give HCI a new experimental variable: when an interlocutor's knowledge ends.

\begin{acks}
Hiromu Yakura was supported in part by the Japan Science and Technology Agency through the PRESTO Grant no: JPMJPR246B. Robin Schimmelpfennig acknowledges support from the Swiss National Science Foundation Grant no: 100018\_230330.
\end{acks}

\bibliographystyle{ACM-Reference-Format}
\bibliography{refs}

\appendix

\section{Alternative Prediction-Similarity Metrics and Response Examples}
\label{app:prediction-similarity}

We conducted an exploratory robustness analysis of the prediction task using two continuous
text-similarity measures. The analysis used the same 240-participant sample as the main analysis
(121 in the 1930-AI condition and 119 in the Modern-AI condition), comprising 960
participant--item predictions and 2,220 attempts. For the embedding measure, we joined each
participant continuation to the canonical continuation of the AI assigned to that condition and
calculated their cosine similarity using \texttt{text-embedding-3-small}. We removed the shared
sentence stem from both texts before embedding so that fixed prompt wording could not contribute
directly to similarity. Each deployed item had one canonical reference, and cosine distance is one
minus the reported similarity. As a robustness check, we repeated the metric-specific-best
comparison with the original full sentences, including the common stem.

For the likelihood measure, we calculated the negative log-likelihood (NLL) of each participant's
continuation under \texttt{talkie-1930-13b-base}, conditioned on the exact sentence stem. For
continuation tokens $y_1,\ldots,y_T$, the score was the arithmetic mean of
$-\log p_{\theta}(y_t\mid s,y_{<t})$ across $t=1,\ldots,T$. The average is over continuation
tokens only, including the natural leading-space boundary after the stem; lower values indicate
that the continuation is more compatible with the base model.
The stem therefore affects the conditional probabilities but contributes no tokens to the loss
average. In a 12-item diagnostic, mismatching continuations and stems increased mean NLL per token
from 4.605 to 5.373, in the expected direction.
This is deliberately a common historical-model yardstick. It is \emph{not} likelihood under the
instruction-tuned Talkie system deployed in the 1930-AI condition or under GPT-5.5 in the
Modern-AI condition, and therefore should not be interpreted as a symmetric comparison of how
well participants predicted their assigned model.

For each alternative measure, we selected the best attempt on each item (maximum cosine
similarity or minimum mean token NLL), averaged across the participant's four items, and compared
conditions with two-sided Welch tests. The two metric-specific tests were Holm-corrected. We also
repeated the analyses using the attempt selected by the deployed 0--5 rubric, attempt 1 only, and
the final submitted attempt (Table~\ref{tab:prediction-similarity}). On the metric-specific best
attempts, participants in the Modern-AI condition were closer to their assigned reference in
embedding space than participants in the 1930-AI condition ($M=0.531$, $SD=0.078$ vs. $M=0.458$,
$SD=0.071$; 1930-AI minus Modern-AI $=-0.073$, 95\% CI $[-0.092,-0.054]$,
$t(234.63)=-7.57$, $p_{\mathrm{adj}}<.001$, $d=-0.98$). This reproduces the direction of the
rubric-based result with a continuous semantic-proximity measure. The full-sentence robustness
analysis produced the same direction with the expected higher absolute cosine values
($M=0.825$ vs. $M=0.809$; difference $=-0.016$, 95\% CI $[-0.025,-0.008]$,
$p<.001$).

Under the common base-Talkie yardstick, the 1930-AI condition instead had lower mean token NLL
than the Modern-AI condition ($M=3.631$, $SD=0.825$ vs. $M=4.103$, $SD=0.630$; difference
$=-0.472$, 95\% CI $[-0.659,-0.286]$, $t(224.25)=-4.99$,
$p_{\mathrm{adj}}<.001$, $d=-0.64$). Thus, participants exposed to the historical system wrote
continuations that were more compatible with the historical base model. Relative to the assigned
canonical responses scored under that same base model, the metric-specific human responses were
1.18 NLL per token higher in the 1930-AI condition and 0.09 higher in the Modern-AI condition. These
reference gaps are descriptive because the canonical responses came from different generators
and themselves differed in base-model likelihood.

\begin{figure}[t]
  \centering
  \includegraphics[width=\linewidth]{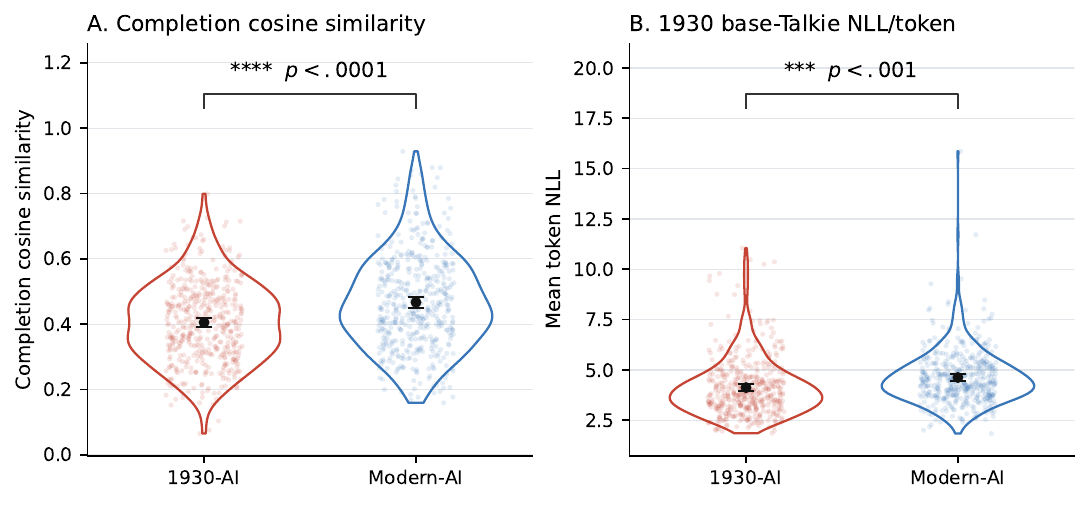}
  \caption{Alternative prediction-similarity measures on first attempts. Light points show all
  first participant--item responses (1930-AI: $n=484$; Modern-AI: $n=476$), and outlines show
  their response-level distributions. Black points and error bars show means and 95\% confidence
  intervals calculated from participant averages ($n=121$ and $n=119$); the tests likewise use
  participants as the independent units and are Holm-corrected across the two metrics. Panel A
  shows completion-only cosine similarity to the assigned AI's canonical continuation, where
  higher values indicate greater proximity (1930-AI: $M=0.405$, $SD=0.082$; Modern-AI:
  $M=0.467$, $SD=0.095$; 1930-AI minus Modern-AI $=-0.062$, 95\% CI
  $[-0.084,-0.039]$, $t(231.33)=-5.37$, $p_{\mathrm{adj}}<.0001$, $d=-0.69$). Panel B shows
  mean token NLL under the common 1930 base-Talkie model, where lower values indicate greater
  historical-model compatibility (1930-AI: $M=4.123$, $SD=1.054$; Modern-AI: $M=4.622$,
  $SD=0.940$; difference $=-0.500$, 95\% CI $[-0.753,-0.246]$, $t(235.76)=-3.88$,
  $p_{\mathrm{adj}}=.000138$, $d=-0.50$). This NLL is not likelihood under the deployed
  instruction-tuned Talkie or GPT-5.5. Significance markers denote *** $p<.001$ and
  **** $p<.0001$.}
  \Description{Two violin plots show all first participant--item responses in the 1930-AI and Modern-AI conditions. Modern-AI has higher completion-only cosine similarity to its assigned canonical continuation, while 1930-AI has lower negative log-likelihood under the common 1930 base-Talkie model. Black intervals show uncertainty calculated from participant averages.}
  \label{fig:prediction-similarity-conditions}
\end{figure}

\begin{figure}[t]
  \centering
  \includegraphics[width=\linewidth]{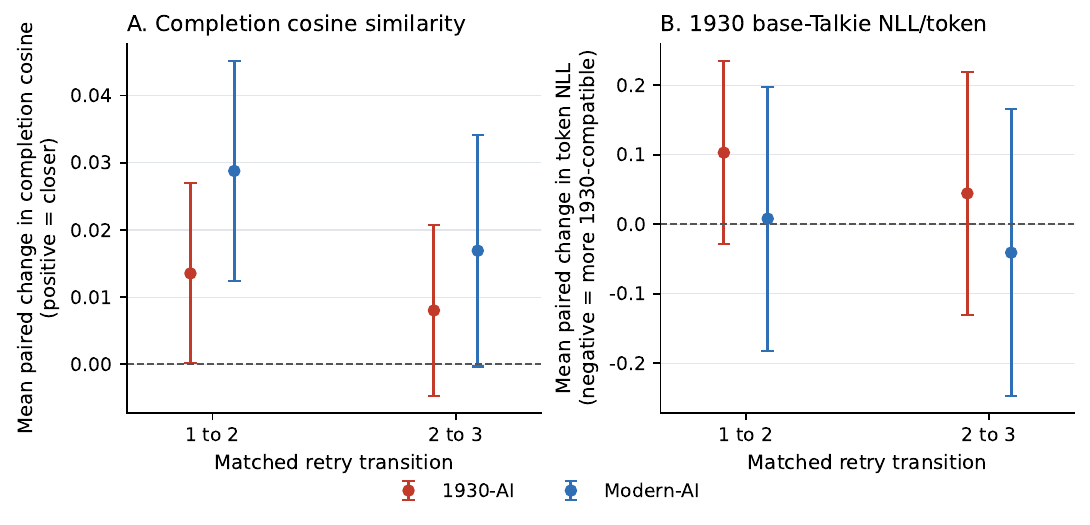}
  \caption{Matched changes across retry transitions. Points show the mean within-participant
  change after first calculating changes for the same participant--item pair; error bars show
  paired 95\% $t$ confidence intervals across participants. In Panel A, completion cosine increased
  from attempt 1 to 2 by $.014$ in 1930-AI (95\% CI $[.000,.027]$; 374 pairs from 120
  participants) and by $.029$ in Modern-AI (95\% CI $[.012,.045]$; 329 pairs from 115
  participants). From attempt 2 to 3, the changes were $.008$ (95\% CI $[-.005,.021]$; 305 pairs
  from 119 participants) and $.017$ (95\% CI $[-.000,.034]$; 252 pairs from 111 participants),
  respectively. In Panel B, the corresponding mean token-NLL changes were $.103$ (95\% CI
  $[-.029,.235]$) and $.008$ (95\% CI $[-.182,.198]$) from attempt 1 to 2, and $.044$ (95\% CI
  $[-.130,.219]$) and $-.041$ (95\% CI $[-.247,.166]$) from attempt 2 to 3; all NLL intervals
  included zero. Positive cosine change means closer to the assigned canonical continuation;
  negative NLL change means greater 1930 base-Talkie compatibility. Red denotes 1930-AI and blue
  denotes Modern-AI. These estimates apply only to retried items.}
  \Description{Two point-and-interval plots show matched changes from attempt 1 to 2 and attempt 2 to 3. Completion cosine generally increases among retried items, especially at the first retry, whereas all confidence intervals for changes in negative log-likelihood include zero.}
  \label{fig:prediction-similarity-attempts}
\end{figure}

The condition difference in completion cosine remained in the rubric-selected, first-attempt,
and final-attempt analyses. Figures~\ref{fig:prediction-similarity-conditions}
and~\ref{fig:prediction-similarity-attempts} separate the initial condition comparison from the
observed change across attempts. Mixed-effects models of all attempts, with participant random
slopes and item random intercepts, found no condition-by-attempt interaction for completion
cosine ($b=-0.0030$, $p=.588$) or for base-model NLL ($b=0.104$, $p=.080$). Thus, neither
alternative measure provided reliable evidence that the condition difference changed across
successive attempts. At the attempt level, the rubric score correlated moderately with completion
cosine ($\rho=.547$, $p<.001$) but only weakly with base-model NLL ($\rho=-.112$, $p<.001$),
further showing that the two measures capture different aspects of imitation.

\begin{table*}[t]
\caption{Alternative measures of prediction similarity. Values are participant-level means of four item scores, with standard deviations in parentheses. Difference is 1930-AI minus Modern-AI. The headline cosine excludes the shared sentence stem; the full-sentence version is reported as a robustness check. Higher cosine similarity indicates greater proximity to the assigned reference; lower NLL indicates greater compatibility with the 1930 base Talkie model. The two metric-specific-best tests use Holm-adjusted $p$ values; sensitivity analyses show unadjusted exploratory $p$ values.}
\Description{Table comparing alternative prediction-similarity measures between AI conditions across several attempt-selection rules. Modern-AI responses have higher cosine similarity to their assigned reference, whereas 1930-AI responses have lower negative log-likelihood under the common 1930 base-Talkie model. The pattern is consistent across sensitivity analyses.}
\label{tab:prediction-similarity}
\centering\scriptsize
\setlength{\tabcolsep}{3pt}
\begin{tabular}{@{}llcccc@{}}
\toprule
Selection & Measure & 1930-AI $M$ ($SD$) & Modern-AI $M$ ($SD$) & Difference [95\% CI] & $p$ \\
\midrule
Metric-specific best & Completion-only cosine & $0.458$ ($0.071$) & $0.531$ ($0.078$) & $-0.073$ [$-0.092$, $-0.054$] & $<.001$ \\
 & Base-Talkie NLL/token & $3.631$ ($0.825$) & $4.103$ ($0.630$) & $-0.472$ [$-0.659$, $-0.286$] & $<.001$ \\
\addlinespace
Rubric-selected best & Completion-only cosine & $0.431$ ($0.082$) & $0.507$ ($0.087$) & $-0.076$ [$-0.098$, $-0.055$] & $<.001$ \\
 & Base-Talkie NLL/token & $4.099$ ($1.024$) & $4.583$ ($0.847$) & $-0.484$ [$-0.722$, $-0.245$] & $<.001$ \\
\addlinespace
First attempt & Completion-only cosine & $0.405$ ($0.082$) & $0.467$ ($0.095$) & $-0.062$ [$-0.084$, $-0.039$] & $<.001$ \\
 & Base-Talkie NLL/token & $4.123$ ($1.054$) & $4.622$ ($0.940$) & $-0.500$ [$-0.753$, $-0.246$] & $<.001$ \\
\addlinespace
Final attempt & Completion-only cosine & $0.420$ ($0.083$) & $0.495$ ($0.096$) & $-0.075$ [$-0.098$, $-0.052$] & $<.001$ \\
 & Base-Talkie NLL/token & $4.214$ ($1.045$) & $4.601$ ($0.824$) & $-0.387$ [$-0.627$, $-0.148$] & $.002$ \\
\addlinespace
Full-sentence robustness & Full-sentence cosine & $0.809$ ($0.030$) & $0.825$ ($0.035$) & $-0.016$ [$-0.025$, $-0.008$] & $<.001$ \\
\bottomrule
\end{tabular}
\end{table*}

Figure~\ref{fig:prediction-similarity-conditions} visualizes all first responses, before feedback
could affect a participant's completion. The points and distribution outlines use the 960
participant--item responses. To avoid treating the four responses from each participant as
independent for inference, the black means and confidence intervals and the condition tests are
calculated after averaging each participant's four first responses.

Figure~\ref{fig:prediction-similarity-attempts} instead isolates matched changes among
participant--item pairs that were actually retried. We first calculated each item-level change
and then averaged these changes within participant, so participants remain the independent units
for the confidence intervals. These estimates describe retried items only and are not estimates
of change for the full sample.

Table~\ref{tab:prediction-examples} provides verbatim first-attempt predictions from the 1930-AI
condition across the observed performance range. To avoid selecting examples for rhetorical fit,
we ranked first attempts by the mean of their completion-embedding-closeness percentile and inverse-NLL
percentile, selected observations nearest the 90th, 50th, and 10th percentiles, and required
distinct participants and items. Participant identifiers are not shown.

\begin{table*}[t]
\caption{Examples of first-attempt predictions in the 1930-AI condition. Examples were selected deterministically near the 90th, 50th, and 10th percentiles of the mean of completion-only embedding- and NLL-based closeness percentiles, with distinct participants and items. Higher cosine similarity and lower mean token NLL indicate greater proximity. Human text is reproduced verbatim, and these sentences do not necessarily reflect the authors' view.}
\Description{Three examples of first-attempt predictions from the 1930-AI condition, selected from high, median, and low similarity bands. Each row shows the sentence stem, the participant's verbatim prediction, the 1930-AI reference response, cosine similarity, and negative log-likelihood.}
\label{tab:prediction-examples}
\centering\scriptsize
\setlength{\tabcolsep}{4pt}
\begin{tabularx}{\linewidth}{@{}p{0.06\linewidth}p{0.17\linewidth}>{\raggedright\arraybackslash}X>{\raggedright\arraybackslash}Xcc@{}}
\toprule
Band & Stem & Human first attempt & 1930-AI reference & Completion cosine & NLL/token \\
\midrule
High & The death penalty for a person who has committed murder is & fair, because it is to punish the murderer. & justifiable, because murder is a crime so dreadful that it cannot be too severely punished. & 0.664 & 2.958 \\
Median & Striking or spanking a child who misbehaves is & not a good practice. & justifiable, because it prevents him from doing worse. & 0.248 & 2.575 \\
Low & Marriage between a man and woman of different races is & is bad because of tainted bloodline & unsuccessful, because differences of race are hostile to harmony in marriage. & 0.275 & 7.284 \\
\bottomrule
\end{tabularx}
\end{table*}

\section{Details of the materials used on the study}
\label{app:materials}

This appendix expands on the three stages we took to prepare the prediction task items and the scoring mechanism of the LLM judge, which were summarized in \secref{sec:case:materials}.

\subsection{Item selection}
\label{app:selection}
 
Items were drawn from two sources of repeated survey questions bearing on moral topics, each entered by a documented rule. The first is a pool of about 420 candidate wordings: the corpus of repeated survey items assembled by \citet{mastroianni2023illusion}, together with vintage Gallup World Poll (GWP)~\cite{tortora2010gallup} questions first fielded between 1935 and 1969, taken from Gallup's published archive with their original wording, year, and marginal. Most items in the corpus are \emph{change} questions (i.e., whether some behavior is more common than ten years ago) which a model cannot answer, so the archival questions carry most of this pool. The second is the General Social Survey (GSS)~\citep{davern2025gss}, whose relevant items begin in 1972. Because the GSS is far larger than can be screened by reading, candidates were ranked before screening: each item was scored on the absolute change in the proportion agreeing between the earliest wave in which it appears and the most recent, subject categories were ranked by their median change, and the largest-moving item in each category was carried forward, giving a ranked pool of 44.
 
Candidates from both pools were then screened on the same criteria, applied to the verbatim source question rather than to its topic, since the period marginal attaches to the question and not to the subject matter, and a morally loaded topic is not the same thing as a morally loaded question.
 
\begin{enumerate}[label=(\roman*), leftmargin=*, itemsep=1pt, topsep=3pt]
\item \emph{Answerable by a pre-1930 model.} The question must not turn on anything the model cannot know. GWP 1954 item asks respondents to approve or disapprove of a Supreme Court decision handed down that year. Where the practice at issue predates the referent, the question is recast as that practice---here, teaching children of different races in the same schools---and its marginal is then declared one step removed from the stem shown. Where it does not, the item is dropped.
\item \emph{An evaluation, not a fact, a report, or a prediction.} A morally loaded topic can fail in several ways. ``Do you think cigarette smoking is one of the causes of lung cancer?'' asks for a factual belief, so movement on it records epidemiology rather than morality. ``Did you, yourself, happen to attend church or synagogue in the last seven days?'' asks for a behavioral self-report: the marginal is an attendance rate, not a judgment that attending is commendable. ``Would government lotteries produce an unwholesome gambling spirit?'' asks for a prediction about consequences. And a question asking whether something is better or worse than it was ten years ago asks for a perceived trend, which is a judgment about change rather than about the practice.
\item \emph{No explicit moral vocabulary in the stem.} The evaluative word must come from the completion, not from the prompt. GWP 1969 item asks whether it is \emph{wrong} for a man and a woman to have relations before marriage; the stem names only the act, i.e., \stem{For a man and a woman to have relations before they are married is}, and leaves the verdict to the blank.
\item \emph{Intelligible in both eras.} The stem must name something a reader in 1930 and a reader today would both recognize, so that neither is answering a different question. GWP 1936 item asks about ``the birth control movement'', an organized campaign of its day that a present-day participant would not place; the stem states the practice instead, \stem{For married couples to use means to limit the number of their children is}. Where no such restatement is available, the item is dropped.
\item \emph{Sustained by the model in the completion frame.} Checked on live draws before deployment: the model must return an evaluation, not a description or a measurement. A stem leading with the amount a government spends, for instance, draws a size adjective (\stem{large}) rather than a verdict.
\item \emph{Ethically clearable} under the approved protocol.
\end{enumerate}

Finally, two authors independently rated each surviving candidate for moral load, which was judged on the verbatim question, not the topic, and for pre-1930 compatibility, with disagreements resolved by discussion.
This screening left 41 items, twelve were selected (eight from the GWP and four from the GSS). Each participant saw four of the twelve, drawn at random, preceded by one practice item that is excluded from all analyses.
 
The twelve selected items concern married women's paid work, contraception, capital punishment, the physical discipline of children, euthanasia, abortion, interracial marriage, premarital relations, same-sex relations, women in high political office, police permits for gun purchase, and the employment of an openly homosexual teacher. The two models take opposite sides on ten of the twelve. The set is a sample of contested moral positions rather than a neutral sample of moral topics, and performance on the prediction task should not be read as an estimate over topics in general.
 
The period marginals were not used to select items, stems or references; each reference was chosen on the model's own majority direction (\appref{app:modelresponseselection}). Compared afterward as a check on historical consistency, the direction agrees on five of the eight GWP-derived items, and on five of seven if abortion is set aside as not strictly comparable, its 1962 marginal being case-specific where the stem is general. Because no survey from the 1930s exists for these topics (GWP begins in 1935, the GSS in 1972), the oldest available marginal stands in for the model's own period, so agreement is evidence about the model's \emph{voice} rather than a test against opinion from its own era (\vreq{1}).
 
\subsection{Reformulation into stems}
\label{app:reformulation}
 
The source wording was frozen as the provenance record, and the participant-facing text derived from it followed a common rule: drop the interrogative frame, preserve the act being evaluated, introduce no new vocabulary, and end the sentence immediately before the evaluative completion. Two to five candidate stems were written per item and tested against the model at temperature~0 for grammatical completion, preservation of the source proposition, and intelligibility in both eras. Three needed iteration: ``For a woman to end a pregnancy is'' was read as childbirth, so ``deliberately end a pregnancy'' was adopted; public-relief stems leading with the amount spent drew a quantity adjective rather than a verdict; a married-women's-work variant with a gerund subject made the model complete with a verb under every prompt tried, localizing the defect to the stem. Three of the four GSS stems retain the survey's own substantive phrase; \texttt{FEPOL}, an agree--disagree statement, was expressed as the practice at issue. The GSS stems are adaptations rather than derivations under the stricter rule applied to the GWP items.
 
\subsection{Selecting the model responses}
\label{app:modelresponseselection}
 
The historical model has no instruction tuning and cannot be instructed: an instruction to avoid circular reasoning left its own words in the completions while the circularity persisted, so the fix had to be structural. Continuing the sentence with ``because'' made the model restate its verdict as its reason; continuing with \stem{. The reason is that} did not. The adopted protocol, used for both models, presents the stem in a first call after \emph{Complete the sentence with a judgment in one word}, then inserts the verdict into a second call, \emph{Answer in at most twelve words}, continuing \stem{$\langle$stem$\rangle$ $\langle$verdict$\rangle$. The reason is that}.
 
Each item was sampled 25 times at temperature~1.0 plus one greedy draw, used as a reproducible anchor and not as the item's stance. Reasons essentially never repeat, so there is no modal \emph{completion}, only a modal \emph{verdict}; one reference per item was selected by reading against eight criteria in fixed precedence, with coherence, non-emptiness, and non-circularity absolute, and majority direction, present-day intelligibility, and \textit{guessability} outranking brevity. The stance a completion takes was never itself a criterion, and no text was edited: every deployed reference is a verbatim draw. Euthanasia and physical discipline are two-sided and declared bimodal. Two departures apply to the modern model and make the arms non-identical: it would not take a side (94 of 300 baseline draws returned a verdict about the question rather than a position on it), so its first call was constrained by a two-value polarity schema; and because its draws converge on one verdict word per item, we fixed polarity and grounds at their modal values and drew the verdict \emph{word} from the model's attested vocabulary for that item under a no-repetition constraint across the set.
 
\subsection{Scoring}
\label{app:scoring}
 
A prediction receives 5 when it takes the same side as the reference and gives substantively the same grounds; 4 for closely related grounds; 3 for the same side with different or missing grounds; 2 for a responsive answer taking no side; 1 for the opposite side; 0 for a response that does not answer the moral question. Fluency, period style, and modern vocabulary are never rewarded or penalized in themselves. The rubric was developed before any participant data in two rounds of three-rater annotation, the second on 60 predictions from five simulated participant strategies: exact agreement 58\%, all raters within one scale point on every row, Krippendorff's $\alpha=.872$, ICC(2,1) $=.912$, with six written decision rules raising exact agreement on the disputed rows to 73\%. The deployed judge (\texttt{gpt-5.4-mini}) applied the frozen rubric to one participant--reference pair per call, returning the stance category, the rule applied, and an integer score. Participants' answers were capped at the reference length plus five words.

\section{Supplementary Figures}

\begin{figure}[h!]
  \centering
  \includegraphics[width=\linewidth]{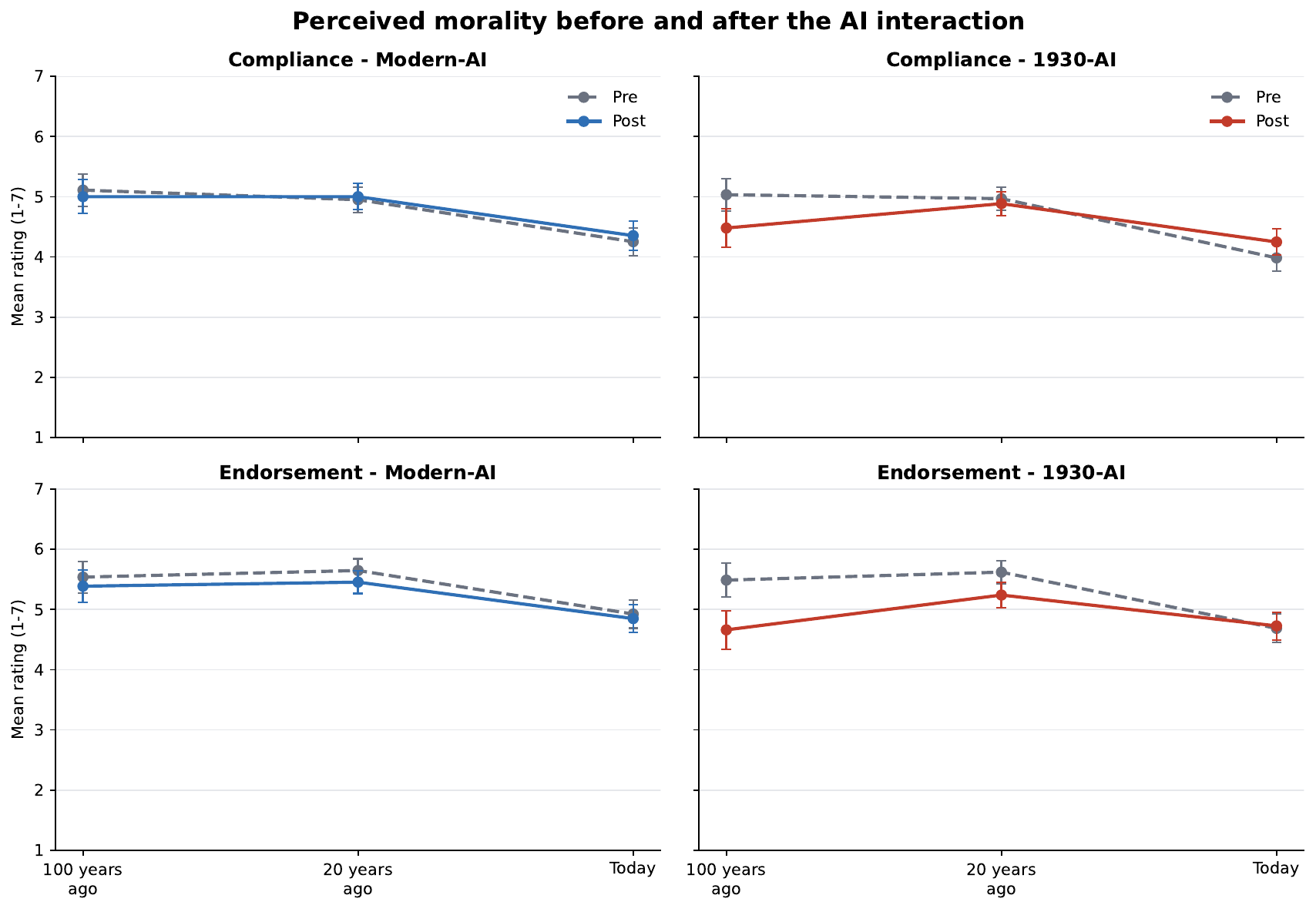}
  \caption{\textbf{Perceived morality before and after the AI interaction.}
  Mean ratings of moral compliance (top row) and moral-value endorsement (bottom row) for people 100 years ago, 20 years ago, and today, shown separately for the Modern-AI (left column) and 1930-AI (right column) conditions.
  Dashed gray lines represent pre-interaction ratings, and solid colored lines represent post-interaction ratings.
  Points indicate condition means and error bars show 95\% confidence intervals.}
  \Description{Four-panel line plots of perceived moral compliance before and after the AI interaction. Ratings are shown for people 100 years ago, 20 years ago, and today in each AI condition. The largest pre-to-post change occurs for the 100-year rating in the 1930-AI condition.}
  \label{fig:moral-trajectories}
\end{figure}

\end{document}